\documentclass[%
aps, %
prb, 
twocolumn, superscriptaddress, 
showpacs, %
a4paper,
amsmath, amssymb]{revtex4}
\usepackage{color}
\usepackage{graphicx}
\usepackage{subfigure}
\usepackage{bm}
\usepackage{braket}
\usepackage{mathrsfs}
\usepackage[normalem]{ulem}
\usepackage{makeidx}
\usepackage{hhline}
\usepackage[
	dvipdfmx, %
	bookmarks=true, %
	bookmarksnumbered=true, %
	bookmarkstype=toc, %
	breaklinks=true, %
	colorlinks=true, %
	linkcolor=blue, %
	citecolor=blue, %
	urlcolor=blue, %
	pdfstartview={FitH} %
]{hyperref}
\usepackage[capitalise]{cleveref} 
\crefname{section}{Sec.}{Secs.}
\crefname{subsection}{Sec.}{Secs.}
\newcommand{\ii}{i}

\newcommand{\dd}[1]{d #1\,}
\renewcommand{\Im}{\mathrm{Im}\,}

\newcommand{\round}[2]{\frac{\partial#1}{\partial#2}}
\newcommand{\sign}[1]{\mathrm{sign}(#1)}
\begin{document}
\title{Transport Properties of Dirac Ferromagnet}
\date{\today}
\author{Junji Fujimoto}
\email[E-mail address: ]{jfujimoto@blade.mp.es.osaka-u.ac.jp}
\affiliation{Department of Materials Engineering Science, Osaka University, Machikaneyama 1-3, Toyonaka, Osaka 560-8531 Japan}
\author{Hiroshi Kohno}
\email[E-mail address: ]{kohno@s.phys.nagoya-u.ac.jp}
\affiliation{Department of Physics, Nagoya University, Furo-cho, Chikusa-ku, Nagoya, 464-8601, Japan}
\begin{abstract}
 We propose a model ferromagnet based on the Dirac Hamiltonian in three spatial dimensions, and study its transport properties which include  anisotropic magnetoresistance (AMR) and anomalous Hall (AH) effect.
 This relativistic extension allows two kinds of ferromagnetic order parameters, denoted by $\bm{M}$ and $\bm{S}$, which are distinguished by the relative sign between the positive- and negative-energy states (at zero momentum) and become degenerate in the non-relativistic limit. 
 Because of the relativistic coupling between the spin and the orbital motion, both $\bm{M}$ and $\bm{S}$ induce anisotropic deformations of the energy dispersion (and the Fermi surfaces) but in mutually opposite ways. 
 The AMR is determined primarily by the anisotropy of the Fermi surface (group velocity), and secondarily by the anisotropy of the damping; the latter becomes important for ${\bm M}=\pm{\bm S}$, where the Fermi surfaces are isotropic. 
 Even when the chemical potential lies in the gap, the AH conductivity is found to take a finite non-quantized value, $\sigma_{ij} = -(\alpha /3\pi^2 \hbar ) \epsilon_{ijk} S_k $, where $\alpha$ is the (effective) fine structure constant. 
 This offers an example of Hall insulator in three spatial dimensions. 
\end{abstract}
\pacs{72.25.-b, 71.15.Rf}
\maketitle
\section{Introduction}
\label{sec:intro}
 Recently, spintronics is an active area of research with fundamental as well as applicational interest.
 In spintronic phenomena based on ferromagnetic materials, one of the central interests is the interaction between electric current and magnetization, such as giant magnetoresistance and electrical manipulation of magnetization. 
 While these phenomena do not require spin-orbit coupling (SOC) in principle, they are expected to acquire new features in the presence of SOC. 
It is thus important to study the effects of SOC on various phenomena in spintronics.  

 As one of the simplest models of ferromagnetism containing SOC, we propose in this paper a Stoner-like model based on the Dirac Hamiltonian in three spatial dimensions (3D). 
 Such models of ferromagnetism (with relativistic effects) have been considered previously in two different ways. 
 One, introduced by MacDonald and Vosko,\cite{MacDonaldVosko} is characterized by a ferromagnetic order parameter (which we represent by $\bm{M}$ in this paper) having opposite signs between the positive- and negative-energy states. 
 The other, proposed by Ramana and Rajagopal,\cite{Ramana1981} is described by an order parameter (which we denote by $\bm{S}$) having the same sign in both states. 
 While these models were originally motivated by the interest in true relativistic effects (``true'' Dirac model) in the context of {\it ab initio} calculations,\cite{MacDonaldVosko, Ramana1981, Crepieux2001, Ebert2011}
such models may also find applications as low-energy effective models (``effective'' Dirac model) of electrons in solids.
  Especially in the latter case, in view of the fact that the ferromagnetism arises as a spontaneous symmetry breaking governed by electron interaction, there seems to be no reason to exclude either of the two order parameters {\it a priori}.
 Based on this observation, we propose a model which contains both $\bm{M}$ and $\bm{S}$ in general.
 In the following, we call this model a ``Dirac ferromagnet'' for brevity. 

 As for the ``effective'' Dirac model, it is known that electrons in some crystals, such as bismuth\cite{Wolf1964} and bulk states of 3D topological insulators,\cite{Zhang2009} are effectively described by Dirac-type Hamiltonians derived from the $k \cdot p$ perturbation theory.
 If such materials become ferromagnetic, for example by doping magnetic impurities, they will be described as a Dirac ferromagnet. 
 Candidate materials are Mn-doped Bi$_2$(Se,Te)$_3$\cite{Hor2010,Checkelsky2012} 
and Cr-doped (Bi,Sb)$_2$Te$_3$.\cite{Zhang2013,Samarth}

 In this paper, we study magneto-transport properties of the Dirac ferromagnet in its homogeneous state, which include anisotropic magnetoresistance (AMR)\cite{Thomson1857, Potter1974, McGuire1975, Kokado2012} and anomalous Hall effect (AHE)\cite{Hall1881, Kaplus1954, Nagaosa2010}.
 While we present a general formulation by retaining both $\bm{M}$ and $\bm{S}$, explicit results will be demonstrated mostly for three typical cases; 
(i) $\bm{S} = {\bm 0}$,
(ii) $\bm{M} = {\bm 0}$, and 
(iii) $\bm{M} = \bm{S}$. 
 Two factors have been identified that determine the AMR, the anisotropy of the Fermi surface (group velocity) and the anisotropy of the damping. 
 In general, the former effect is much stronger than the latter, but in case ${\bm M}= {\bm S}$, where the Fermi surfaces are isotropic, AMR is determined by the latter.  
 As for AHE, we found a new type of Hall insulator state in which the Hall conductivity is finite and proportional to $\bm{S}$ (hence not quantized) while the longitudinal conductivity vanishes.
 This may explain the peculiar behavior of Cr-doped (Bi,Sb)$_2$Te$_3$ found recently. \cite{Samarth}

 This paper is organized as follows. 
 We define the model in \cref{sec:model} and calculate the conductivity tensor in \cref{sec:calculation}.
 The main results are presented in \cref{sec:resultsanddiscussion}, where AMR and the anomalous Hall (AH) conductivity are shown for the three typical cases, (i)--(iii), and the key factors are discussed. 
 Summary is given in \cref{sec:summary}. 
 Some details of the calculation are presented in the Appendices. 
 \cref{app:gamma} gives the damping constants explicitly, 
and \cref{app:magnetization,app:spin,app:coexistent} give the calculation of AH conductivity 
for the three cases, respectively. 
 Symmetry properties of the conductivity tensor are studied in \cref{app:sym}.

\section{Model and Green's function}
\label{sec:model}
\subsection{Model}
 We consider an electron system described by a $4 \times 4$ Dirac Hamiltonian with additional two kinds of ferromagnetic order parameters, $\bm{M}$ and $\bm{S}$, 
\begin{equation}
 \mathcal{H}_0 =  \hbar c \bm{k} \cdot \bm{\sigma} \rho_1 + mc^2 \rho_3 - \bm{M} \cdot \bm{\sigma} \rho_3 - \bm{S} \cdot \bm{\sigma}  , 
\label{eq:Hamiltonian}
\end{equation}
and subject to impurity potentials, 
\begin{equation}
 V_{\rm imp} ({\bm r}) = u \sum_i \delta ({\bm r} -{\bm R}_i)
\label{eq:Vimp}
.\end{equation}
Here $\bm{\sigma} = (\sigma^{x}, \sigma^{y}, \sigma^{z})$ are the Pauli matrices in spin space,  $\rho_i \, (i = 1, 2, 3)$ are the Pauli matrices in electron-positron (particle-hole) space, and $m$, $c$ and $\bm{k}$ are the mass, velocity and wave vector, respectively, of a Dirac particle. 
 The total Hamiltonian is given by 
\begin{equation}
 H_{\rm tot} = \sum_{\bm k} \psi^\dagger_{\bm k} ({\cal H}_0 - \mu ) \psi_{\bm k} + \int d{\bm r} \psi^\dagger ({\bm r}) V_{\rm imp} ({\bm r}) \psi ({\bm r})
\label{eq:H_tot}
,\end{equation}
where $\psi ({\bm r})$ is a four-component Dirac spinor field, $\psi_{\bm k}$ is its Fourier transform, and  $\mu$ is the chemical potential. 
 We treat the impurity potential $V_{\rm imp}$ perturbatively in the Born approximation (see the next subsection). 
 In the following calculations, we put $c = \hbar = 1$ and recover them in the results. 

 Because of the $\rho_3$ matrix, ${\bm M}$ acts oppositely between the positive- and negative-energy states 
(at ${\bm k}={\bm 0}$),\cite{com1} whereas ${\bm S}$ acts with the same sign.
 In the ``true'' Dirac model, ${\bm M}$ physically represents ``magnetization'' and couples to real magnetic fields whereas ${\bm S}$ represents ``spin'', which would not couple to any physical fields in the microscopic Hamiltonian.\cite{com5,com6}
 Note, however, that this does not mean that ${\bm S}$ is unphysical and unsuitable for an order parameter as suggested in Ref.~\onlinecite{Crepieux2001}. 
 This is because a ferromagnetism arises as a spontaneous symmetry breaking, which is governed by the content of the interaction.\cite{com2}  

 When the ferromagnetism is driven by magnetic doping in a solid whose low-energy effective Hamiltonian is of Dirac-type (``effective'' Dirac model), \cite{Wolf1964, Zhang2009} the resulting order parameter will be ${\bm S}$ (${\bm M}$) if the exchange interaction of Dirac particles with the magnetic impurity have the same (opposite) sign between the positive-energy state (conduction band) and negative-energy state (valence band).\cite{com_elmag_eff} 

\begin{figure*}[t]
 \subfigure[]{\includegraphics[width=0.3\textwidth]{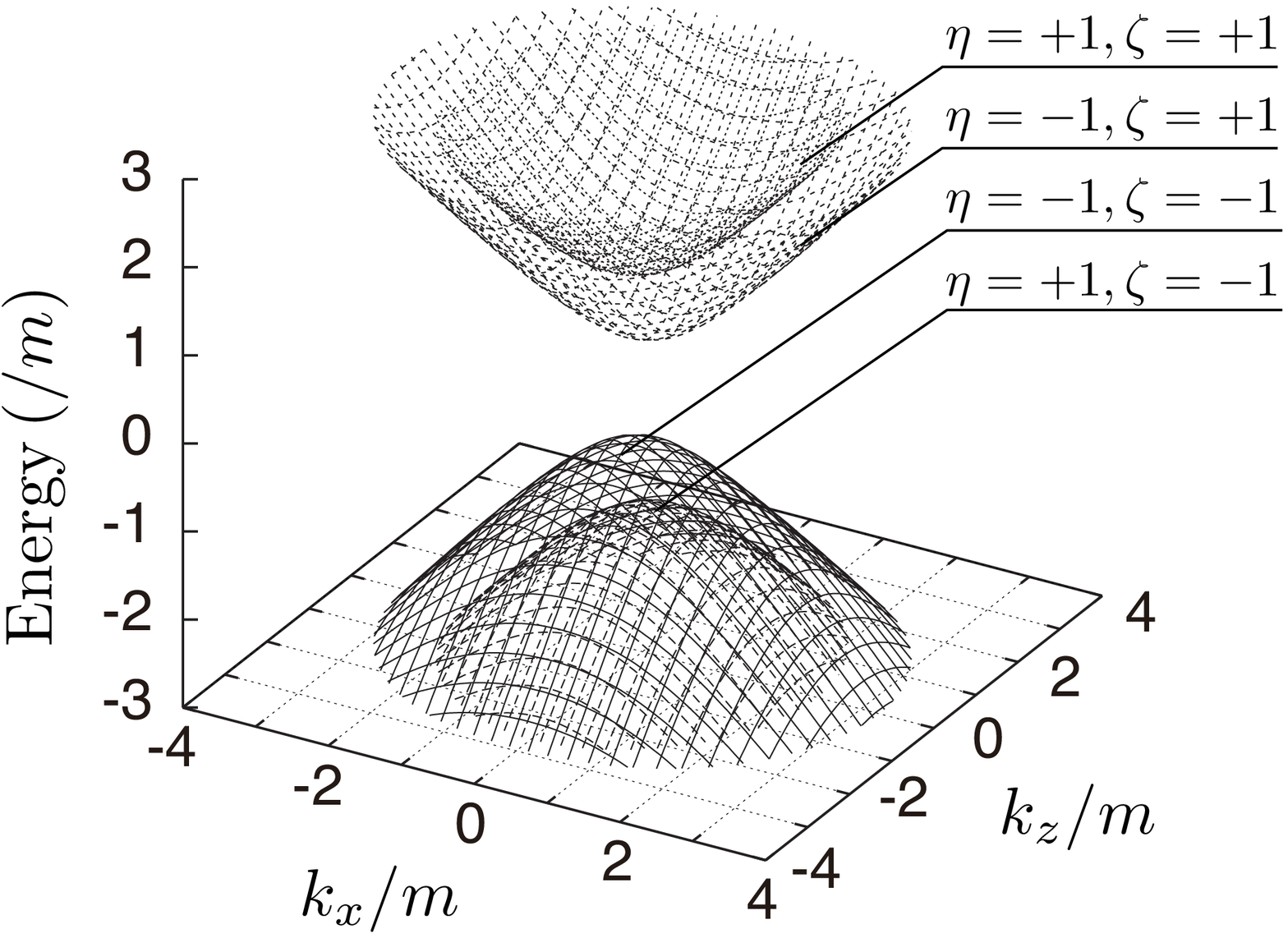}\label{fig:sub:dispersion}}
 \subfigure[]{\includegraphics[width=0.6\textwidth]{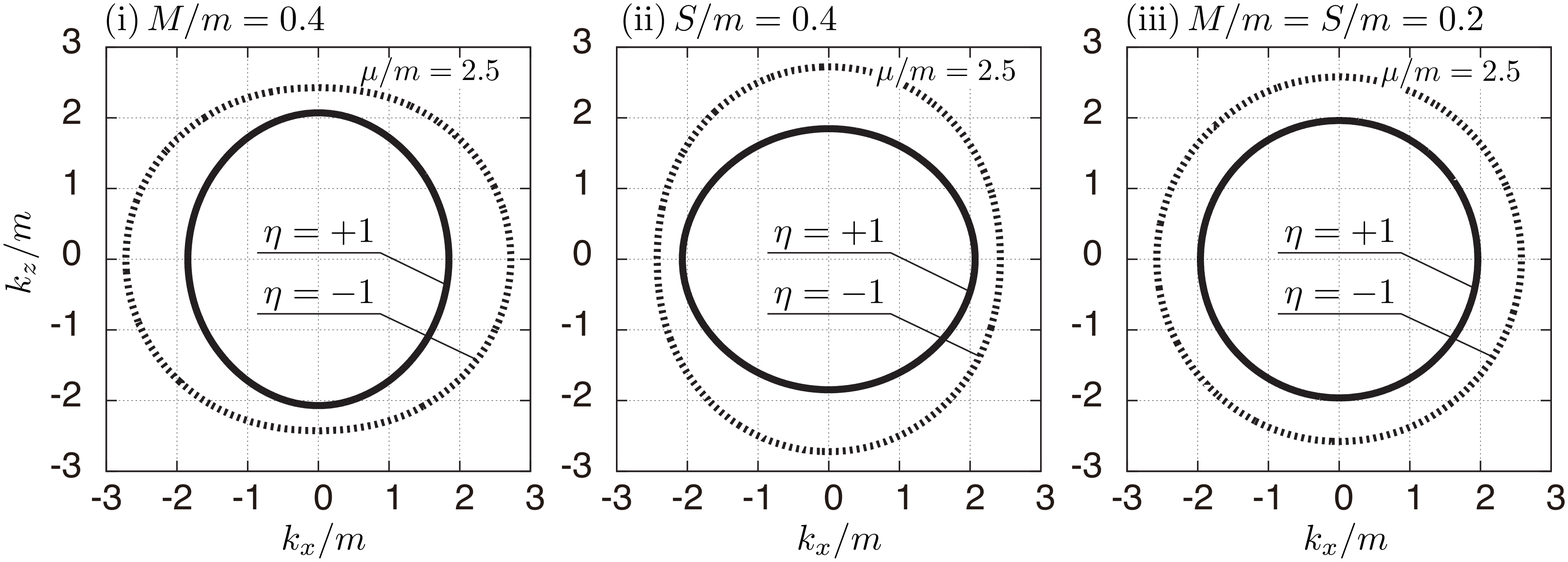}\label{fig:sub:fermi-surfaces}}
 \caption%
{Energy dispersion and Fermi surfaces for $\bm{M}=M\hat{z}$ and $\bm{S}=S\hat{z}$.
\subref{fig:sub:dispersion} Energy dispersion for $M=0.4m, S=0$ (case (i)) plotted against $k_z$ and $k_x$ (or $k_y$).
 The $\eta = \pm1$ specifies spin, and $\zeta = \pm 1$ specifies upper/lower Dirac bands (positive/negative energy states). 
\subref{fig:sub:fermi-surfaces} Fermi surfaces for the three typical cases, (i)--(iii).} 
 \label{fig:dispersion-fermi_surface}
\end{figure*}

 In this paper, we assume that the order parameters are given. 
 We restrict ourselves to the case that $\bm{M}$ and $\bm{S}$ are uniform and mutually parallel, and take the  $z$-axis along their direction, ${\bm M} = M \hat z$, ${\bm S} = S \hat z$. 
 In addition, we assume that $ M + S < m$ to avoid the closing of the original gap (due to $m$); see below. 
 The values of $M$ and $S$ are otherwise arbitrary, but some explicit results will be displayed for the following three typical cases; 
\begin{enumerate}
\item[(i)] $\bm{M} = M \hat{z},\, \bm{S} ={\bm 0}$ \ \ \ (\lq\lq ${\bm M}$ model'')
\item[(ii)] $\bm{M} = {\bm 0} ,\, \bm{S} = S \hat{z}$ \ \ \ (\lq\lq ${\bm S}$ model'')
\item[(iii)] $\bm{M} = \bm{S} = S \hat{z}$ \ \ \ (\lq\lq coexistent model'')
\end{enumerate}
In these special cases, the energy dispersion takes relatively simple forms, 
\begin{align}
& \zeta \sqrt{k^2 + m^2 + M^2 + 2\eta M \sqrt{k_\perp^2 + m^2} }
, \\
& \zeta \sqrt{k^2 + m^2 + S^2 + 2\eta S \sqrt{k_z^2 + m^2} }
, \\
& \zeta \sqrt{k^2 + (m + \eta S)^2 } + \eta S
,\end{align} 
for (i), (ii) and (iii), respectively, where $k_\perp^2 = k_x^2 + k_y^2$. 
 (See \cref{eq:dispersion_M,eq:dispersion_S,eq:dispersion_MS}.) 
 Here, $\zeta = \pm 1$ specifies positive/negative energy states, and $\eta = \pm 1$ specifies spin states. 
 \Cref{fig:sub:dispersion} shows the energy dispersion for case (i).
 The Fermi surfaces at $\mu / m = 2.5$ are shown in \cref{fig:sub:fermi-surfaces} for the three cases.
 In contrast to ordinary (non-relativistic) ferromagnets, such as the Stoner model, Fermi surfaces are deformed in cases (i) and (ii) by the presence of ferromagnetic order parameters and become anisotropic.
 The anisotropy is opposite between the two Fermi surfaces in each case, and between the two cases (i) and (ii).
 Such anisotropic deformation of the Fermi surfaces due to ferromagnetism and SOC has been noted in ferromagnets with Dirac-,\cite{Crepieux2001} Luttinger-\cite{Nguyen2007} and Rashba-type\cite{Wang2010} SOC. 
 In case (iii), the Fermi surfaces remain isotropic. 
%
\subsection{Green's function}
The unperturbed Green's function $G^{(0)}_{\bm{k}} (\epsilon) = ( \epsilon - \mathcal{H}_{0} )^{-1}$ can be expressed as
\begin{equation}
G^{(0)}_{\bm{k}} (\epsilon) = \frac{1}{D_{\bm{k}} (\epsilon)} \sum_{ \substack{ \mu = 0,1,2,3 \\ \nu = 0,x,y,z } } g^{(0)}_{\mu \nu} (\epsilon) \rho_{\mu} \sigma^{\nu}
\label{eq:G0}
,\end{equation}
where $\sigma^0$ and $\rho_0$ are unit matrices. 
 We have defined 
\begin{align}
D_{\bm{k}} (\epsilon)
	& = ( \epsilon^2 - k^2 - m^2 - S^2 + M^2)^2
\notag \\ & \hspace{2em}
 	- 4 \{ (\epsilon M + m S)^2 + (S^2 - M^2) k_z^2 \}
\\	& =
	\left\{ \epsilon^2 - \epsilon_k^2 - 2 \Delta_{\bm{k}}(\epsilon) \right\}
	\left\{ \epsilon^2 - \epsilon_k^2 + 2 \Delta_{\bm{k}}(\epsilon) \right\}
\label{eq:Dk_general}
, \\
 \epsilon_k
	& = \sqrt{ k^2 + m^2 + S^2 - M^2 } \label{eq:E_k}
, \\
\Delta_{\bm{k}} (\epsilon)
	& = \sqrt{ \Omega^2 + (S^2-M^2) k_z^2 }
, \\
\Omega
	& = \epsilon M + m S
,\end{align}
and $g^{(0)}_{\mu \nu} (\epsilon)$'s are listed in \cref{tab:g0_general}.
\begin{table*}[t]
\begin{center}
\caption{The coefficients $g^{(0)}_{\mu \nu}$ of the unperturbed Green's function $G^{(0)}_{\bm k} (\epsilon)$ 
for $\bm{M} = M \hat{z}$ and $\bm{S} = S \hat{z}$. }
\begin{tabular}{|c||c|c|c|c|}
\hline
$(\mu, \nu)$ & 0 & $x$ & $y$ & $z$ \\ \hhline{|=#=|=|=|=|}
0 	& $ \epsilon(\epsilon^2 - \epsilon_k^2) - 2 M \Omega $
	& $ - 2 S k_z k_x $
	& $ - 2 S k_z k_y $
	& $ - S ( \epsilon^2 - \epsilon_k^2 + 2 k_z^2) - 2 m \Omega $
\\ \hline
1	& $ - 2 (\epsilon S + m M) k_z $
	& $ (\epsilon^2 - \epsilon_k^2) k_x $
	& $ (\epsilon^2 - \epsilon_k^2) k_y $
	& $ k_z (\epsilon^2 - \epsilon_k^2 + 2 S^2 - 2 M^2) $
\\ \hline
2	& $ 0 $
	& $ - 2 \Omega k_y $
	& $ 2 \Omega k_x $
	& $ 0 $
\\ \hline
3	& $ m (\epsilon^2 - \epsilon_k^2) + 2 S \Omega $
	& $ 2 M k_z k_x $
	& $ 2 M k_z k_y $
	& $ M(\epsilon^2 - \epsilon_k^2 + 2k_z^2) - 2 \epsilon \Omega $
\\ \hline
\end{tabular}
\label{tab:g0_general}
\end{center}
\end{table*}

 We evaluate the self-energy due to impurity scattering in the first Born approximation, 
\begin{equation}
 \Sigma^{\rm R} (\epsilon) = n_{\mathrm{i}} u^2 \sum_{\bm{k}} G^{(0)}_{\bm{k}} (\epsilon + \ii 0)
\label{eq:self-energy}
,\end{equation}
where $n_{\mathrm{i}}$ is the concentration of impurities. 
 Neglecting the real part, we consider only the imaginary part, which is calculated as 
\begin{equation}
\Im \Sigma^{\rm R} (\epsilon) = - \sum_{\mu, \nu} \gamma_{\mu \nu} (\epsilon) \rho_{\mu} \sigma^{\nu}
\label{eq:dampingparameter}
,\end{equation}
with
\begin{align}
\gamma_{\mu \nu} (\epsilon) = - n_{\mathrm{i}} u^2 \sum_{\bm{k}} \Im \frac{1}{D_{\bm{k}} (\epsilon + \ii 0)} g^{(0)}_{\mu \nu} (\epsilon)
\label{eq:gamma}
.\end{align}
 Only the four components, $(\mu, \nu) = (0, 0)$, $(0, z)$, $(3, 0)$ and $(3, z)$, are finite and others vanish by symmetry. 
 (See \cref{app:gamma} for explicit results.) 
 The renormalized (retarded) Green's function is expressed as
\begin{equation}
\begin{split}
\{G^{\rm R}_{\bm{k}} (\epsilon) \}^{-1}
& = \epsilon - \mathcal{H}_0 - \ii \, \mathrm{Im} \Sigma^{\rm R} (\epsilon)
\\ & = 
	\epsilon + \ii \gamma_{0 0}
	- \rho_1 \bm{k} \cdot \bm{\sigma}
	- (m - \ii \gamma_{3 0}) \rho_3
\\ & \hspace{2em}
	+ (M + \ii \gamma_{3 z} ) \rho_3 \sigma^z
	+ (S + \ii \gamma_{0 z} ) \sigma^z
.\end{split}
\label{eq:G}
\end{equation}
 This is obtained from $G^{(0)}_{\bm{k}} (\epsilon)$ by the replacements, 
$\epsilon \to \epsilon + \ii \gamma_{0 0}, m \to m - \ii \gamma_{3 0}, S \to S  + \ii \gamma_{0 z}, M \to M + \ii \gamma_{3 z}$.
As in \cref{eq:G0}, we express it as
\begin{equation}
G^{\mathrm{R}}_{\bm{k}}(\epsilon)
	= \frac{1}{D^{\mathrm{R}}_{\bm{k}}(\epsilon)}
		\sum_{\mu, \nu} g^{\mathrm{R}}_{\mu \nu} (\epsilon) \rho_{\mu} \sigma^{\nu}
\notag
.\end{equation}

 Let us write the denominator $D_{\bm{k}}^{\rm R} (\epsilon)$ as
\begin{equation}
 D_{\bm{k}}^{\rm R} (\epsilon) = D^{\prime} + \ii D^{\prime\prime}
,\end{equation}
with the real ($D'$) and imaginary ($D''$) parts, which are given by 
\begin{align}
D^{\prime}
	& = D_{\bm{k}}
		+ \mathcal{O} (\gamma^2)
\label{eq:D'}
, \\
D^{\prime\prime}
	& = 4 \{ (\epsilon^2 - \epsilon_k^2) \Gamma_1 -  2 \Omega \Gamma_2 - 2 k_z^2 \Gamma_3 \}
		+ \mathcal{O} (\gamma^3) 
\label{eq:D''}
, \\
\Gamma_1
	& = \epsilon \gamma_{0 0} + m \gamma_{3 0} - S \gamma_{0 z} + M \gamma_{3 z}
, \\
\Gamma_2
	& = M \gamma_{0 0} - S \gamma_{3 0} + m \gamma_{0 z} + \epsilon \gamma_{3 z}
, \\
\Gamma_3
	& = S \gamma_{0 z} - M \gamma_{3 z}
.\end{align}
 In this paper, we assume that the effects of impurity scattering are weak, and calculate the conductivity tensor in the leading order with respect to the damping parameters $\gamma_{\mu \nu}$, which are collectively denoted as $\gamma$ in the following as well as in \cref{eq:D',eq:D''}, where terms of 
$\mathcal{O}(\gamma^2)$ and $\mathcal{O}(\gamma^3)$, respectively, or higher are suppressed. 
%
\section{Calculation of Conductivity}
\label{sec:calculation}
 The dc conductivity tensor  $ \sigma_{i j} $ $(i, j = x,y,z)$ is given by the Kubo formula\cite{Kubo1957} as 
\begin{equation}
\sigma_{i j} = \lim_{\omega \to 0} \frac{Q_{i j}^{\rm R}(\omega) - Q_{i j}^{\rm R} (0)}{\ii \omega}
\label{eq:conductivity}
,\end{equation}
where $Q_{i j}^{\rm R}(\omega)$ is the current-current retarded response function. 
 In this paper, we neglect vertex corrections and evaluate it from 
\begin{align} 
Q_{i j} (\ii \omega_{\lambda})
	& = - e^2 T \sum_{n} \sum_{\bm{k}}
		\mathrm{tr} \left[
			v_{i} \tilde G_{\bm{k}}(\ii \epsilon_{n} + \ii \omega_{\lambda})
			v_{j} \tilde G_{\bm{k}}( \ii \epsilon_{n})
		\right] 
\label{eq:Q}
,\end{align}
via the analytic continuation, $\ii \omega_\lambda \to \omega + i0$, 
where $\epsilon_{n} = (2n+1)\pi T$ and $\omega_{\lambda} = 2 \lambda \pi T$ are Matsubara frequencies, 
and 
\begin{equation}
 {\bm v} = \rho_1 {\bm \sigma}  
\end{equation}
is the velocity matrix. 
 In \cref{eq:Q}, we have defined 
\begin{align}
 \tilde G_{\bm{k}}( \ii \epsilon_{n})  &= (\ii \epsilon_{n} + \mu - {\cal H}_0 - \Sigma (\ii \epsilon_{n}))^{-1} 
\label{eq:G_tilde}
,\end{align}
where $\Sigma (\ii \epsilon_{n})$ and $\Sigma^{\rm R} (\epsilon)$ [\cref{eq:self-energy}] are mutually related via the analytic continuation. 
 (This $\tilde G_{\bm{k}}( \ii \epsilon_{n})$ differs from $G_{\bm{k}}( \ii \epsilon_{n})$ only in the presence of $\mu$.)
 The temperature $T$ is assumed to be zero, $T=0$, throughout. 
 The result is written as 
\begin{equation}
\sigma_{i j} = \sigma^{(1)}_{i j} + \sigma^{(2)}_{i j} + \sigma^{(3)}_{i j}
,\end{equation}
with 
\begin{widetext}
\begin{align}
\sigma_{i j}^{(1)}
	& = - \frac{e^2}{4 \pi} \sum_{\bm{k}} \mathrm{tr} \left[
			v_{i} (G_{\bm{k}}^{\rm R}(\epsilon) - G_{\bm{k}}^{\rm A}(\epsilon))
			v_{j} (G_{\bm{k}}^{\rm R}(\epsilon) - G_{\bm{k}}^{\rm A}(\epsilon))
		\right] \bigl. \Bigr|_{\epsilon=\mu}
\label{eq:AMR}
, \\
\sigma_{i j}^{(2)}
	& = \frac{e^2}{4 \pi} \sum_{\bm{k}} \mathrm{tr} \left[
			v_{i} G_{\bm{k}}^{\rm R}(\epsilon)
			v_{j} G_{\bm{k}}^{\rm A}(\epsilon)
			- v_{i} G_{\bm{k}}^{\rm A}(\epsilon)
			v_{j} G_{\bm{k}}^{\rm R} (\epsilon)
		\right] \bigl. \Bigr|_{\epsilon=\mu}
\label{eq:AHE1}
, \\
\sigma_{i j}^{(3)}
	& = - \frac{e^2}{4 \pi} \sum_{\bm{k}} \int_{-\infty}^{\mu} \dd{\epsilon}
	\lim_{\epsilon' \rightarrow \epsilon} (\partial_{\epsilon} - \partial_{\epsilon'})
	\, \mathrm{tr} \left[
		v_{i} G^{\rm R}_{\bm{k}}(\epsilon)
		v_{j} G^{\rm R}_{\bm{k}} (\epsilon')
		- v_{i} G^{\rm A}_{\bm{k}} (\epsilon)
		v_{j} G^{\rm A}_{\bm{k}} (\epsilon')
	\right] 
\label{eq:AHE2}
.\end{align}
\end{widetext}
 Here, $\sigma_{i j}^{(1)}$ is symmetric with respect to $i$ and $j$ and contributes to the longitudinal conductivity, whereas $\sigma_{i j}^{(2)}$ and $\sigma^{(3)}_{i j}$ are anti-symmetric and contribute to the Hall conductivity. 
 Because of the remaining $\epsilon$-integral, $\sigma^{(3)}_{i j}$ is often called a ``Fermi-sea term'', whereas $\sigma_{i j}^{(1)}$ and $\sigma_{i j}^{(2)}$ are called ``Fermi-surface terms''.\cite{Streda1982}
 In the present model, only the diagonal components are finite for $\sigma_{i j}^{(1)}$ by symmetry. 

 As stated above, we calculate $\sigma_{i j}$ in the leading order with respect to the damping parameters $\gamma_{\mu \nu}$, which are ${\cal O}(\gamma^{-1})$ for the longitudinal conductivity, and ${\cal O}(\gamma^0)$ for the Hall conductivity.
\subsection{AMR}
\begin{figure*}[htbp]
 \begin{center}
  \includegraphics[width=0.5\textwidth]{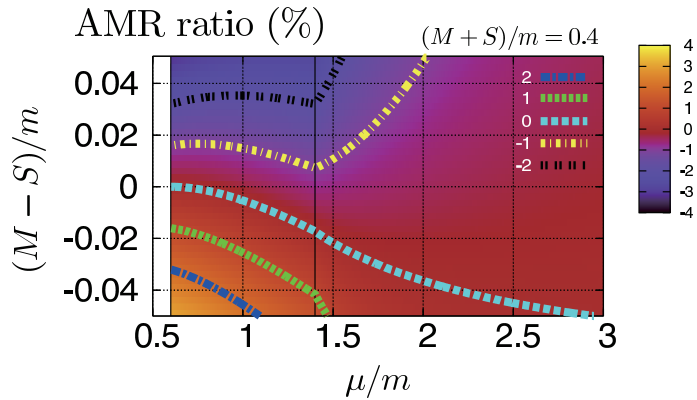}
 \end{center}
 \caption{(Color online) The AMR ratio $(\sigma_{\perp} - \sigma_{\parallel}) /(\sigma_{\perp} + \sigma_{\parallel}) \times 100$ plotted in the plane of $\mu$ and $M - S$, with $M +S = 0.4m$ kept constant.
 The white region ($\mu / m < 0.6$) represents the energy gap. 
 There is a single Fermi surface for $0.6 < \mu / m < 1.4$, and are two Fermi surfaces for $\mu /m > 1.4$.}
 \label{fig:AMR-ratio_2}
\end{figure*}

 We first consider the symmetric part, $\sigma^{(1)}_{i j}$.
 Putting $\sigma_{\perp} \equiv \sigma^{(1)}_{xx}$ $(=\sigma^{(1)}_{yy})$ and $\sigma_{\parallel} \equiv \sigma^{(1)}_{zz}$, the difference $\Delta \sigma = \sigma_{\perp} - \sigma_{\parallel}$ defines AMR.
 Substituting the Green's function \cref{eq:G} in the form of \cref{eq:G0} into \cref{eq:AMR}, dropping the damping in the numerator as $ g^{\rm R (A)}_{\mu \nu} (\epsilon) \simeq g^{(0)}_{\mu \nu} (\epsilon)$ since the leading contributions with respect to $\gamma$ are of our interest, taking the traces of $\rho_{\mu}$ and $\sigma^{\nu}$ matrices, and using
\begin{equation}
\left( \frac{1}{D^{\rm R}_{\bm{k}}} - \frac{1}{D^{\rm A}_{\bm{k}}} \right)^2
 = - 4 \left[ \frac{ D^{\prime \prime} }{ ( D^{\prime} )^2 + ( D^{\prime \prime} )^2 } \right]^2
 \simeq - \frac{2 \pi}{| D^{\prime \prime} |} \delta ( D^{\prime} )
,\end{equation}
$\sigma^{(1)}_{i j}$ is expressed as 
\begin{equation}
\sigma_{i j}^{(1)}
 = 2 e^2 \sum_{\bm{k}}
  \left( C^{(1)} \delta_{i j} + C_{i j}^{(2)}  \right)
  \frac{\delta (D' ) }{| D^{\prime \prime} |}
   \bigl. \Biggr|_{\epsilon=\mu}
\label{eq:sigma_10}
,\end{equation}
where $\delta_{i j}$ is the Kronecker's delta, and 
\begin{align}
C_{}^{(1)}
	& = \sum_{\mu, \nu} s_{\mu} \eta_{\nu} g^{(0)}_{\mu \nu} (\epsilon) g^{(0)}_{\mu \nu} (\epsilon) 
, \\
C_{i j}^{(2)}
	& =  \sum_{\mu} 2 s_{\mu} g^{(0)}_{\mu i} (\epsilon) g^{(0)}_{\mu j} (\epsilon)
,\end{align}
with
\begin{align}
s_{\mu} & = \left\{
	\begin{array}{c l}
		1
	&	\quad (\mu = 0, 1)
	\\	-1
	&	\quad (\mu = 2, 3)
	,\end{array}
\right.
\\
\eta_{\nu} & = \left\{
	\begin{array}{c c}
	1
	&	\quad (\nu = 0)
	\\	-1
	&	\quad (\nu = x, y, z)
	.\end{array}
\right.
\label{eq:Tr_variable}
\end{align}
 Explicitly,  $C^{(1)}$ and $C_{ij}^{(2)}$'s are given by
\begin{align}
C^{(1)}
	& = (\epsilon^2 - \epsilon_k^2) D_{\bm{k}} + 8 k_{\perp}^2 \Omega^2 
, \\
C_{\perp}^{(2)}
	& = k_{\perp}^2 \{ D_{\bm{k}} + 8 k_z^2 (S^2 - M^2) \}
, \\
C_{\parallel}^{(2)}
	& = 2 (S^2 - M^2 + k_z^2) D_{\bm{k}}  + 8 \Omega^2 (2 k_z^2 - k_{\perp}^2 ) 
\notag \\ & \hspace{1em}
		+ 16 k_z^2 (S^2 - M^2) ( \epsilon^2 - k_{\perp}^2 - m^2 ) 
,\end{align}
where $C_{\perp}^{(2)} \equiv (C_{xx}^{(2)} + C_{yy}^{(2)})/2$ and $C_{\parallel}^{(2)} \equiv C_{zz}^{(2)}$. 
 The $\delta$-function is resolved as
\begin{equation}
\delta (D^{\prime})
	\simeq \delta (D_{\bm k})  
	= \sum_{\eta = \pm} \frac{\delta ( k_{\perp}^2 - \alpha_{\eta} ) }{4 \Delta_{\bm{k}} (\epsilon)}
		\Theta_{\eta} ( \epsilon )
\label{eq:delta_function}
,\end{equation}
where
\begin{equation}
 \alpha_{\eta} = \epsilon^2 - k_z^2 - m^2 - S^2 + M^2 - 2 \eta \Delta_{\bm{k}} (\epsilon)
\label{eq:alpha_eta}
,\end{equation}
and the function 
\begin{equation}
\Theta_{\eta} (\epsilon)
 = \left\{
	\begin{array}{c c}
		1
	&	\left( \epsilon < - m - \eta | S - M | \text{\ or \ } \epsilon > m + \eta (S + M) \right)
	\\	0
	&	(\text{otherwise}) 
	\end{array}
\right.
\label{eq:Theta_eta}
\end{equation}
assures $\alpha_{\eta} > 0$. 
 Using $\mu^2 - \epsilon_k^2 = 2 \eta \Delta_{\bm{k}}$ ensured by the $\delta$-function in \cref{eq:delta_function}, with 
\begin{equation}
 \Delta_{\bm{k}} \equiv \Delta_{\bm{k}} (\epsilon = \mu)
,\end{equation}
we obtain
\begin{equation}
\begin{split}
\sigma_{A}
	& = \frac{e^2}{2 ( 2 \pi)^2} \sum_{\eta} \Theta_{\eta} (\mu) \int_{0}^{\xi_{\eta}} \dd{k_z}
  \frac{C_A}{| \eta \Delta_{\bm{k}} \Gamma_1 - \Omega \Gamma_2 - k_z^2 \Gamma_3 |}
,\end{split}
\label{eq:AMR2}
\end{equation}
for $A = \perp, \parallel$, where
\begin{align}
 C_{\perp}
 	& = \alpha_{\eta} \Delta_{\bm{k}}
, \\
C_{\parallel}
	& = 2 k_{z}^2 ( \eta \Delta_{\bm{k}} + S^2 - M^2 )^2 / \Delta_{\bm{k}}
,\end{align}
and $\xi_{\eta} \equiv \xi_{\eta} (\epsilon = \mu) $ with
\begin{equation}
\xi_{\eta} (\epsilon) = \sqrt{ \epsilon^2 - m^2 + S^2 - M^2 - 2 \eta | m M + \epsilon S | }
\label{eq:xi_eta}
.\end{equation}
 The remaining $k_z$-integral is performed numerically. 
 The results, plotted in \cref{fig:AMR-conductivity,fig:AMR-ratio} for the three typical cases and in \cref{fig:AMR-ratio_2} for close neighbors of case (iii), will be discussed in \cref{sec:resultsanddiscussion}.
\subsection{AHE}
 We next look at the anti-symmetric parts, $\sigma^{(2)}_{i j}$ and $\sigma^{(3)}_{i j}$. 
 For the Fermi-surface term, $\sigma^{(2)}_{i j}$, the leading-order contributions are $\mathcal{O} (\gamma^0)$, because both the numerator and the denominator are $\mathcal{O} (\gamma)$. 
 Since there are four kinds of damping parameters [see \cref{eq:detail_gamma}], we cannot evaluate $\sigma^{(2)}_{i j}$ by dropping them; 
their ratio determines the value of $\sigma^{(2)}_{i j}$. 
 Substituting $G_{\bm{k}}^{\rm R(A)}$ into \cref{eq:AHE1} and taking the trace, we obtain 
\begin{equation}
\sigma^{(2)}_{i j}
	= - \frac{4 e^2}{\pi} \sum_{\bm{k}}
		\frac{1}{ | D_{\bm{k}}^{\rm R} (\epsilon) | ^2}
		\sum_{\mu, k } \varepsilon_{i j k } s_{\mu} \mathrm{Im} \left[
			g^{\rm R}_{\mu 0} (\epsilon) g^{\rm A}_{\mu k } (\epsilon)
		\right] \bigl. \Bigr|_{\epsilon=\mu}
\label{eq:AHE1-M}
,\end{equation}
where $\varepsilon_{i j k }$ ($i,j,k=x,y,z$) is the Levi-Civita symbol in 3D, and $\sigma^{(2)}_{i j}$ can be nonzero only for $(i, j) = (x, y)$ or $(y, x)$.
 To the leading order in $\gamma$, we approximate as 
\begin{equation}
\frac{1}{| D^{\rm R}_{\bm{k}} (\epsilon) |^2}
	= \frac{1}{ ( D^{\prime} )^2 + ( D^{\prime \prime} )^2 }
	\simeq \frac{\pi}{| D^{\prime \prime} |} \delta ( D^{\prime} ) 
\label{eq:delta_function_2}
,\end{equation}
and
\begin{align}
\sum_{\mu = 0}^3 s_{\mu} \mathrm{Im} \left[
	g^{\rm R}_{\mu 0} (\epsilon) g^{\rm A}_{\mu z} (\epsilon)
\right]
	= - 8 \eta \Delta_{\bm{k}} C_{x y}
 		+ \mathcal{O} (\gamma^3)
,\end{align}
with
\begin{equation}
\begin{split}
C_{x y} & =
	\gamma_{0 0} ( m \Omega + S k_z^2 + \eta S \Delta_{\bm{k}} )
\\ & \hspace{1em}
	+ \gamma_{3 0} ( \epsilon \Omega - M k_z^2 - \eta M \Delta_{\bm{k}} )
\\ & \hspace{1em}
	+ \gamma_{0 z} ( M \Omega - \eta \epsilon \Delta_{\bm{k}} )
\\ & \hspace{1em}
	- \gamma_{3 z} ( S \Omega + \eta m \Delta_{\bm{k}} )
.\end{split}
\end{equation}
 Here, we have put $D_{\bm{k}} = 0$ because of  $\delta (D^{\prime})$ [\cref{eq:delta_function_2}], and used $\mu^2 - \epsilon_k^2 = 2 \eta \Delta_{\bm{k}}$ from \cref{eq:delta_function}.
Therefore, we obtain
\begin{equation}
\begin{split}
\sigma^{(2)}_{x y}
	& = \frac{e^2}{ ( 2 \pi )^2} \sum_{\eta} \eta \, \Theta_{\eta} (\mu)
		\int_0^{\xi_{\eta}} \dd{k_z} \frac{ C_{x y}  }{| \eta \Delta_{\bm{k}} \Gamma_1 - \Omega \Gamma_2 - k_z^2 \Gamma_3|}
,\end{split}
\label{eq:sigma2xy_DF}
\end{equation}
where $\xi_{\eta}$ is given by \cref{eq:xi_eta} with $\epsilon = \mu$.
 The remaining $k_z$-integral is performed numerically and the results are plotted in 
\cref{fig:AHE} for the three typical cases. 
 This is an extrinsic contribution\cite{Crepieux2001, Nagaosa2010} since it can only be obtained in the limit $\omega/\gamma \to 0$.\cite{Nagaosa2010,com3}

The Fermi-sea term $\sigma^{(3)}_{i j}$, which is also ${\cal O}(\gamma^0)$, can be evaluated by dropping all the damping constants. 
 This is an intrinsic contribution\cite{Crepieux2001, Nagaosa2010} since it survives in the `clean' limit $\gamma/\omega \to 0$.\cite{Nagaosa2010}
 Thus we consider
\begin{widetext}
\begin{equation}
\begin{split}
\sigma_{i j}^{(3)}
	& = \frac{4 e^2}{\pi} \sum_{\bm{k}} \int_{-\infty}^{\mu} \dd{\epsilon} \mathrm{Im} \left[
		\frac{1}{{D^{\rm R}_{\bm{k}}(\epsilon)}^2}
	\right]
	\sum_{ \mu = 0}^3  s_{\mu} 
	\sum_{ k = x, y, z }  \varepsilon_{ijk}
	\Bigl\{
		\left( \partial_{\epsilon} g^{(0)}_{\mu 0}(\epsilon) \right) g^{(0)}_{\mu k }(\epsilon)
		- g^{(0)}_{\mu 0}(\epsilon) \left( \partial_{\epsilon} g^{(0)}_{\mu k }(\epsilon) \right)
	\Bigr\}
.\end{split}
\label{eq:AHE2-M}
\end{equation}
\end{widetext}
 This vanishes unless $(i, j) = (x, y)$ or $(y, x)$ as in the case of $\sigma^{(2)}_{i j}$.
 We write 
\begin{equation}
\begin{split}
\Im \left[
	\frac{1}{{D^{\rm R}_{\bm{k}}(\epsilon)}^2}
\right]
	& = \round{}{ D^{\prime} } \frac{ D^{\prime \prime} }{( D^{\prime} )^2 + ( D^{\prime \prime} )^2 }
\\ & \simeq \pi \, \sign{D^{\prime \prime}} \round{}{ D^{\prime} } \delta ( D^{\prime} ) 
,\end{split}
\end{equation}
and resolve the derivative of the $\delta$-function as
\begin{equation}
\begin{split}
\round{}{ D^{\prime} } \delta ( D^{\prime} ) 
	& = \round{\epsilon}{ D^{\prime} } \round{}{ \epsilon } \delta ( D^{\prime} ) 
\\ & = \frac{1 }{ \partial_{\epsilon} D_{\bm{k}}  }
		\round{}{ \epsilon } \left( \sum_{i} \frac{ \delta ( \epsilon - \epsilon_i ) }{ | \partial_{\epsilon} D_{\bm{k}} |_{ \epsilon = \epsilon_i } |} \right)
,\end{split}
\end{equation}
where $\epsilon = \epsilon_i$ ($i=1,\cdots ,4$) are the roots of $D^{\prime} (\epsilon) \simeq D_{\bm{k}} (\epsilon) = 0$.
 Putting
\begin{equation}
\begin{split}
X (\epsilon)
	& \equiv
		\sum_{\mu = 0}^3 s_{\mu} \Bigl\{
		\left( \partial_{\epsilon} g_{\mu 0}(\epsilon) \right) g_{\mu z}(\epsilon)
			- g_{\mu 0}(\epsilon) \left( \partial_{\epsilon} g_{\mu z}(\epsilon) \right)
		\Bigr\}
\\ & = - S D_{\bm{k}} - 8 S [\Delta_{\bm{k}}(\epsilon)]^2 
		- 4 (S k_z^2 + m \Omega ) (\epsilon^2 - \epsilon_k^2) 
\label{eq:X}
,\end{split}
\end{equation}
and integrating by parts, we obtain\cite{com_calc}
\begin{equation}
\begin{split}
\sigma_{x y}^{(3)}
& = 4 e^2 \sum_{\bm{k}, i}  \Biggl\{ 
	\frac{ X(\epsilon) }{ \partial_{\epsilon} D_{\bm{k}} } \frac{ \delta ( \epsilon - \epsilon_{i} ) }{  \partial_{\epsilon} D_{\bm{k}} } \bigl. \Biggr|_{\epsilon = \mu}
  \\ &  \hspace{4em}
 	- \int_{-\infty}^{\mu} \dd{\epsilon}
		\left[ \frac{\partial}{\partial\epsilon} \left(
		 \frac{ X ( \epsilon) }{ \partial_{\epsilon} D_{\bm{k}} }
		\right)  \right] 
		 \frac{ \delta ( \epsilon - \epsilon_{i} ) }{ \partial_{\epsilon} D_{\bm{k}} }
 	\Biggr\}
.\end{split}
\label{eq:sigma3}
\end{equation}
 (The surface term at $\epsilon = - \infty$ vanishes.)
 In the typical three cases, the integrals in \cref{eq:sigma3} can be performed analytically (see \cref{app:magnetization,app:spin,app:coexistent}), giving 
\begin{subequations}
\begin{align}
& \sigma_{x y}^{(3)} (S = 0)
\notag \\ 
& = - \sign{\mu}  \, \frac{e^2}{4 \pi^2} \sum_{\eta} \eta \, \Theta_{\eta} (\mu) \sqrt{\mu^2 - (m + \eta M)^2 }
\label{eq:sigma3_xy_results_M}
, \\
& \sigma_{x y}^{(3)} (M = 0)
\notag \\ 
& = - \frac{S e^2}{3 \pi^2} - \frac{e^2}{4 \pi^2} \sum_{\eta} \eta \Theta_{\eta} (\mu) \sqrt{(| \mu | - \eta S)^2 - m^2}
\label{eq:sigma3_xy_results_S}
, \\
& \sigma_{x y}^{(3)} (M = S)
	= - \frac{S e^2}{3 \pi^2} \,  
	- \frac{e^2}{12 \pi^2} \sum_{\eta} \eta \, \Theta_{\eta} (\mu) \xi_{\eta}
\notag \\ & \qquad  \times
	\frac{ 2 \, \mathrm{sign} (\mu) (\mu - m - 2 \eta S) + \mu + 2 m + \eta S } {| \mu - \eta S |} 
\label{eq:sigma3_xy_results_MS}
,\end{align}
\end{subequations}
for cases (i), (ii) and (iii), respectively.

\section{Results and Discussion}
\label{sec:resultsanddiscussion}
 In this section, we show the results for (i) $(M,S) = (0.4m, 0)$, (ii) $(M,S) = (0, 0.4m)$, and (iii) $(M,S) = (0.2m, 0.2m)$. 
 These parameter sets give the same exchange shift, $\pm (S+M) = \pm 0.4m$, in the upper Dirac band (positive-energy states). 
 Some subtle features present in the Fermi-sea terms (ultraviolet divergence) are also reported. 
\subsection{AMR}
\begin{figure*}[htbp]
 \begin{center}
  \includegraphics[width=\textwidth]{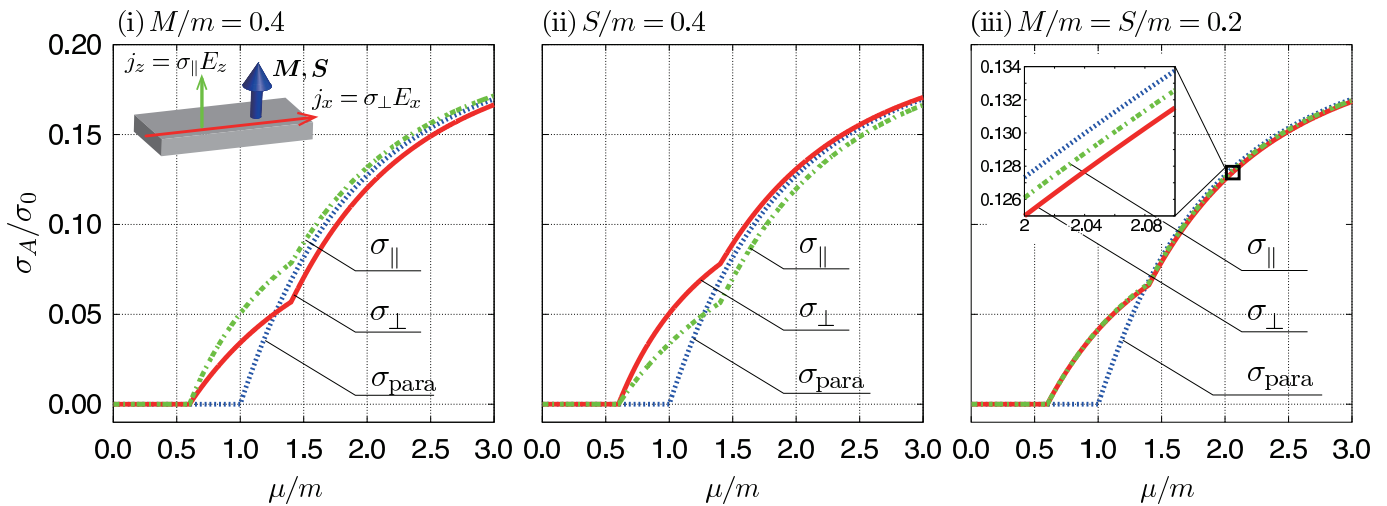}
 \end{center}
\vspace{-3ex}
 \caption%
{(Color online) Diagonal conductivity $\sigma_A$ ($A= \parallel, \perp$)  
as functions of chemical potential $\mu$ for the three typical cases (i)-(iii),  
normalized by $\sigma_0 =  e^2 m^2 c^3 / \hbar^2 \gamma_0$ 
with $\gamma_0 = n_{\mathrm{i}} u^2 m^2 c / \hbar^3$. 
 Those in the paramagnetic state ($M=S=0$) are also plotted as $\sigma_{\rm para}$.
The inset to (i) illustrates mutual directions of $\bm{M}, \bm{S}$ and $\bm{j} = \sigma_A \bm{E}$.
}
 \label{fig:AMR-conductivity}
\end{figure*}

\begin{figure*}[htbp]
 \begin{center}
  \includegraphics[width=\textwidth]{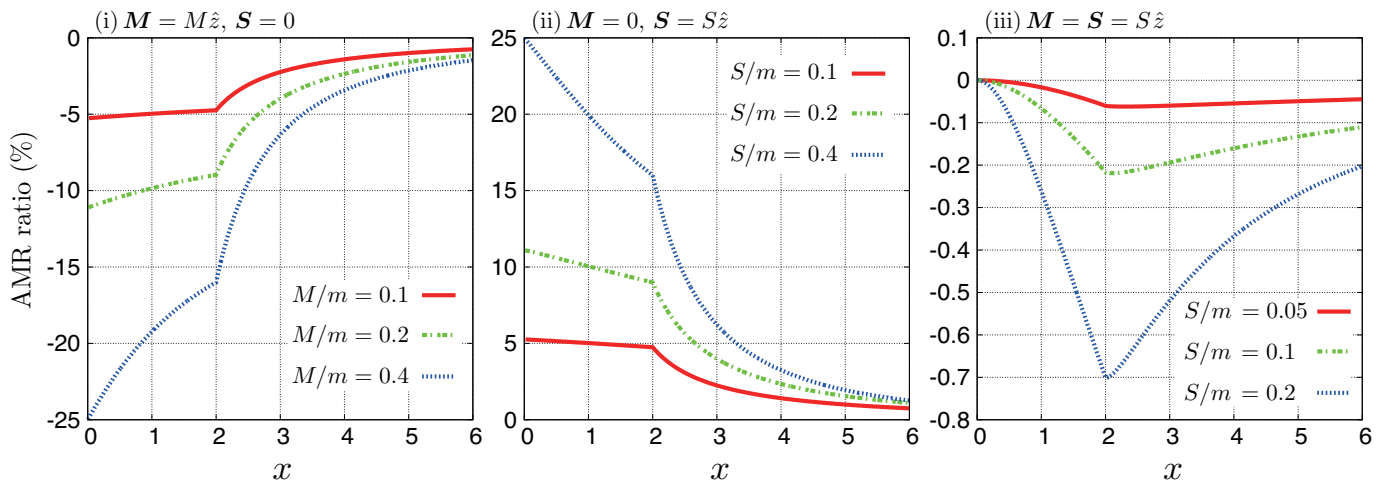}
 \end{center}
 \vspace{2ex}
 \caption{(Color online) The AMR ratio $(\sigma_{\perp} - \sigma_{\parallel}) / (\sigma_{\perp} + \sigma_{\parallel}) \times 100$ plotted against a reduced chemical potential, $x=1+(\mu -m)/(M+S)$, defined by \cref{eq:x}. 
 The bottom of the first (second) band in the upper Dirac band (positive-energy state) corresponds to $x=0$ ($x=2$).}
 \label{fig:AMR-ratio}
\end{figure*}

 The longitudinal conductivity for perpendicular ($\sigma_\perp = \sigma_{xx} = \sigma_{yy}$) or parallel ($\sigma_\parallel = \sigma_{zz}$) configuration, together with the one in the paramagnetic state ($M=S=0$), are plotted in \cref{fig:AMR-conductivity} as functions of $\mu$. 
 The AMR ratio, $(\sigma_\perp - \sigma_\parallel) / (\sigma_\perp + \sigma_\parallel)$, is plotted in \cref{fig:AMR-ratio} against the scaled chemical potential, $x$, defined by 
\begin{equation}
 \mu = m + (M+S)(x-1)
\label{eq:x}
.\end{equation}
 Note that $x=0$ ($x=2$) corresponds to the bottom of the majority- (minority-) spin band in the upper Dirac bands (positive-energy states). 
 As seen, the sign of AMR is opposite between (i) and (ii) and the magnitudes are comparable and large ($ 5 \sim 25\%$). 
 In contrast, for (iii), the magnitude is much smaller ($0.1 \sim 1\%$). 
 To see the physical origin, we note that the (diagonal) conductivity is written as $\sigma_A = e^2 \braket{v^2_{A} \tau_{\bm k}} \nu (\mu)$, where $v_{A}$ is the group velocity in the direction specified by $A$ ($= \perp, \parallel$), $\tau_{\bm k}$ is the relaxation time, $\nu (\mu)$ is the density of states at $\mu$, and $\braket{\cdots}$ represents averaging over the Fermi surfaces. 
 For (iii), since the group velocity is isotropic, the AMR should be ascribed to the anisotropy of $\tau_{\bm k}$. 
 Note that, although the damping constants in the original representation, given by \cref{eq:gamma} [or \crefrange{eq:gamma_appendix_00}{eq:gamma_appendix_3z} for explicit forms], do not depend on ${\bm k}$, the damping of band electrons (obtained after the band diagonalization) depends on ${\bm k}$ in general.
 For (i) and (ii), the two observations made above (magnitude and sign of AMR ratio) indicate that AMR in these cases is totally due to the anisotropy of the band structure (group velocity). 
 (Recall that the deformation of the Fermi surface is opposite between (i) and (ii).)

 The above features can already be seen in the weak-relativistic limit.\cite{Crepieux2001}
 The effective Hamiltonian in this case is derived by the Foldy-Wouthuysen-Tani transformation\cite{FW,Tani} as\cite{com_loose_notation}
\begin{align}
\mathcal{H}_{\rm Pauli}
& =
	\frac{k^2}{2 m}
	- (\bm{M} + \bm{S}) \cdot \bm{\sigma} + V_{\rm imp}
\notag \\ & \hspace{2em}
	+ \frac{1}{2 m^2} \left\{
		(\bm{k} \cdot (\bm{M} - \bm{S}) ) (\bm{k} \cdot \bm{\sigma})
		+ ( \bm{S} \cdot \bm{\sigma} ) k^2
	 \right\}
\notag \\ & \hspace{2em}
	+ \frac{1}{4 m^2} (\bm{k} \times \bm{\nabla} V_{\rm imp}) \cdot \bm{\sigma}
\label{eq:Pauli}
.\end{align}
 The first term in the second line, containing $\bm{M} - \bm{S}$, introduces anisotropy in the energy dispersion. 
 This anisotropy vanishes for $\bm{M} = \bm{S}$; in this case, only the conventional SOC due to the impurity potential connects the spin and the direction of ${\bm k}$, and the AMR for case (iii) can be ascribed to this term.
 The AMR ratio near the region $M=S$ is plotted in \cref{fig:AMR-ratio} as a function of $M - S$ (and $\mu$) while keeping $M + S$ constant.
 The second term in the second line, containing $\bm{S} \cdot \bm{\sigma}$, represents spin-dependent effective mass.\cite{Hirsch1999}
%
\subsection{AHE}
\begin{figure*}[htbp]
 \begin{center}
  \includegraphics[width=\textwidth]{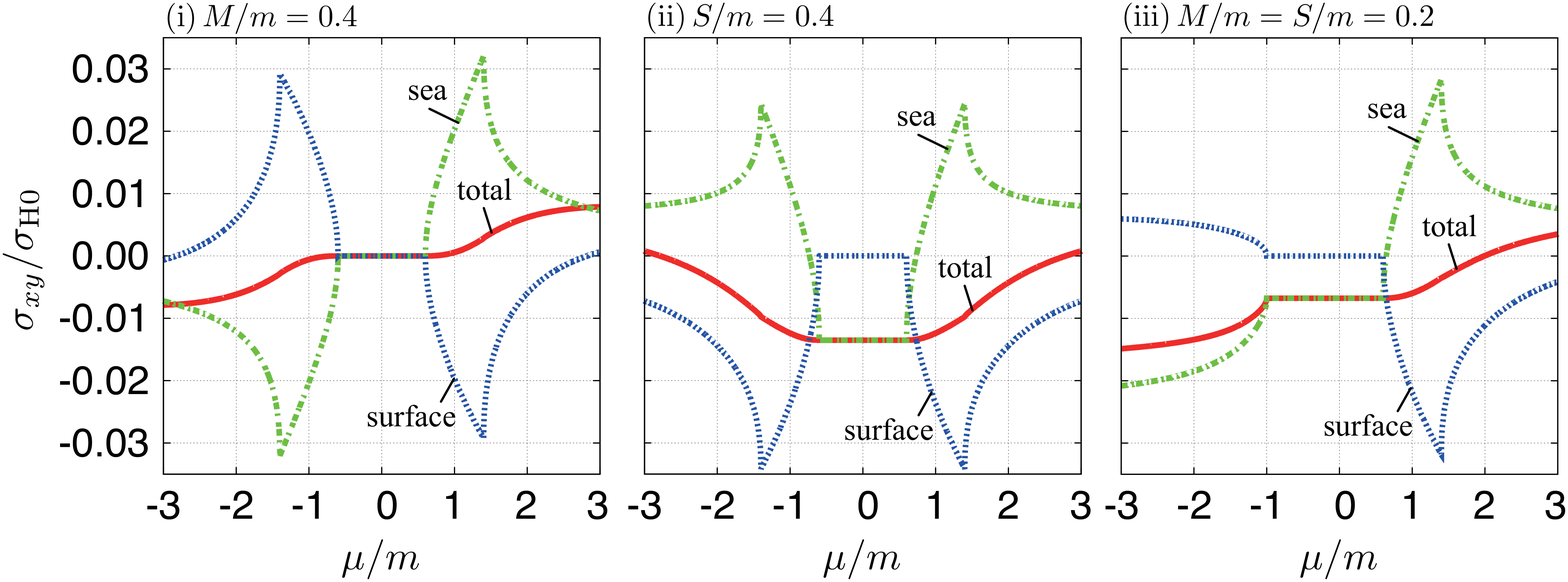}
 \end{center}
\vspace{-3ex}
 \caption%
{(Color online) The off-diagonal (Hall) conductivity $\sigma_{xy} = \sigma^{(2)}_{xy} + \sigma^{(3)}_{xy}$ as functions of $\mu$,
where $\sigma^{(2)}_{xy}$ is the Fermi-surface term and $\sigma^{(3)}_{xy}$ is the Fermi-sea term, each of which are also shown. 
 They are normalized by $\sigma_{\mathrm{H} 0} = e^2 m c/\hbar^2$.}
 \label{fig:AHE}
\end{figure*}

 \Cref{fig:AHE} shows the $\mu$-dependence of $\sigma_{x y}$, which consists of the the Fermi-surface term $\sigma^{(2)}_{xy}$ and the Fermi-sea term $\sigma^{(3)}_{x y}$.
 We see that it is an odd function of $\mu$, $\sigma_{x y} (\mu , M) = - \sigma_{x y}(-\mu , M)$, in case (i), and an even function of $\mu$,  $\sigma_{x y} (\mu , S) = \sigma_{x y} (-\mu , S)$, in case (ii). 
 Indeed, one can show that (see \cref{app:sym}) 
\begin{align}
 \sigma_{x y} (\mu , M, S)
	& = - \sigma_{x y} (-\mu , M, -S)  
\notag \\
	& = \sigma_{x y} (-\mu , -M, S)
\label{eq:symmetry_xy}
,\end{align}
from which the above symmetry properties follow if we put $S=0$ or $M=0$. 

 Remarkably, in cases (ii) and (iii), the AH conductivity takes a finite value, 
\begin{equation}
 \sigma_{x y} (\mu = 0)  = - \frac{e^2 S}{3 \pi^2 \hbar^2 c} = - \frac{\alpha S}{3 \pi^2 \hbar}  , 
\label{eq:Hall_ins}
\end{equation}
even in the insulating state where $\mu$ lies in the band gap.
 Here $\alpha = e^2/\hbar c$ is the (effective) fine structure constant.
 This is obtained from \cref{eq:sigma3_xy_results_S,eq:sigma3_xy_results_MS} by dropping the second terms, which vanish in the gap due to $\Theta_{\eta} (\epsilon)$, while in case (i), it vanishes because of the symmetry, $\sigma_{x y} (\mu =0 , M) = - \sigma_{x y} (\mu =0 , M)$.
 The value is exactly proportional to the ``spin'' order parameter, $S$, hence is not quantized.
 This state thus exemplifies a ``non-quantized Hall insulator'' in three spatial dimensions.

 The finite Hall conductivity at $\mu=0$ (i.e., in the insulating state) arises as interband transitions. 
 For the \lq\lq effective'' Dirac model, this means that the external electric field ${\bm E}$ creates virtual electron-hole pairs and drive them in mutually opposite directions perpendicular to both ${\bm E}$ and ${\bm S}$. 
 This fact (interband transition) can be explicitly demonstrated for the \lq\lq ${\bm S}$ model'', 
where the transitions occur between states with the same $\eta$ and opposite $\zeta$.



 Recently, Samarth {\it et al.} found that in Cr-doped (Bi,Sb)$_2$Te$_3$ the Hall conductivity develops at low temperatures while the resistivity continues to increase down to the lowest temperature.\cite{Samarth}  
These features seem to be consistent with the above Hall insulator state 
with a ferromagnetic order parameter given by ${\bm S}$ 
(purely ${\bm S}$, or a mixture of ${\bm S}$ and ${\bm M}$). 
\begin{figure*}[htbp]
 \begin{center}
  \includegraphics[width=0.35\textwidth]{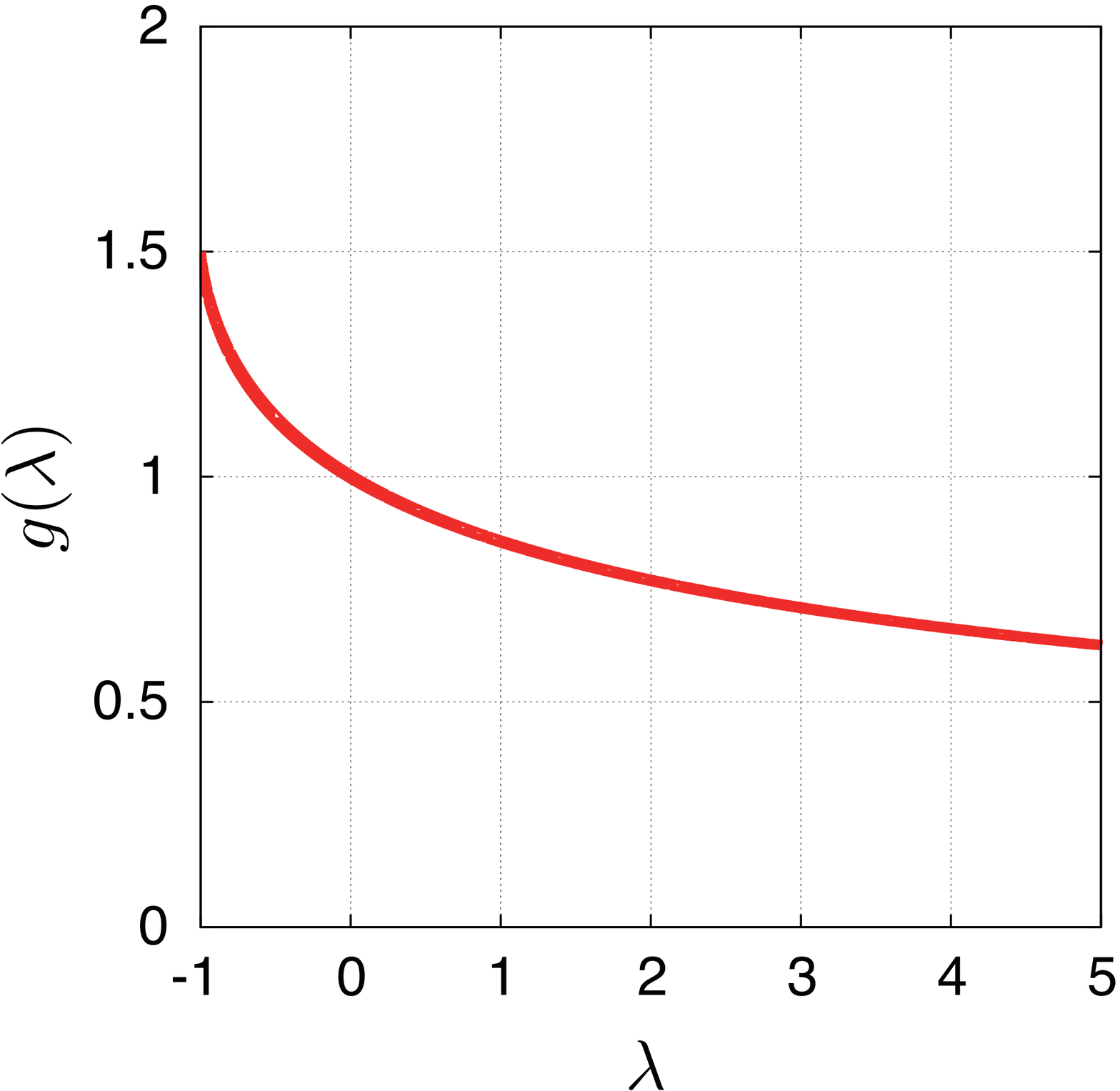}
 \end{center}
 \caption{(Color online) The function $g(\lambda)$ [\cref{eq:g}], where $\lambda$ is 
the ellipticity parameter of the ultraviolet cut-off [\cref{eq:elliptic_cutoff}]. }
 \label{fig:h_delta}
\end{figure*}
%
\subsection{Regularization dependence of Fermi-sea term}
  In cases (ii) and (iii), the Fermi-sea terms contain ultraviolet divergences and we have managed them by introducing a momemtum cut-off $\Lambda$ and letting $\Lambda \to \infty$ at the end. 
 The above results, \cref{eq:Hall_ins} and first terms in \cref{eq:sigma3_xy_results_S,eq:sigma3_xy_results_MS}, were obtained based on the isotropic cut-off, $|{\bm k} | < \Lambda$.
 If other cut-off scheme is used (such as the anisotropic one), one generally obtains a different value. 
 For example, if we take the ``elliptical'' cut-off, 
\begin{equation}
  k_{\perp}^2 + (1 + \lambda) k_z^2 \le \Lambda^2  \quad  (-1 < \lambda < \infty) 
\label{eq:elliptic_cutoff}
,\end{equation}
 \cref{eq:Hall_ins} is replaced by  
\begin{equation}
 \sigma_{xy} (\mu=0) = - \frac{\alpha S}{3 \pi^2 \hbar} \, g (\lambda)
\label{eq:s_xy_g}
,\end{equation}
[so are the first terms of \cref{eq:sigma3_xy_results_S,eq:sigma3_xy_results_MS}], 
where $g(\lambda)$ is given by \cref{eq:g} and plotted in \cref{fig:h_delta}. 
 (See \cref{app:spin,app:coexistent} for details).
 While the isotropic cut-off seems to be most natural, anisotropic cut-off may find its relevance in real materials having crystal anisotropy. 
\section{Summary}
\label{sec:summary}
 We have proposed a model ferromagnet based on the Dirac Hamiltonian in 3D by noting that there are two possible ferromagnetic order parameters, $\bm{M}$ and $\bm{S}$. 
 By restricting ourselves to the case where $\bm{M}$ and $\bm{S}$ are collinear, we have studied its magneto-transport properties, which are AMR and AHE.
 The AMR is found to be determined primarily by the anisotropy of the Fermi surface (group velocity) and secondarily by the anisotropy of the damping. 
 As for AHE, the present model offers an example of non-quantized Hall insulator state in which the Hall conductivity is proportional to $\bm{S}$ while the longitudinal conductivity vanishes.
 We have suggested that this may explain the peculiar property of Cr-doped (Bi,Sb)$_2$Te$_3$ found in a recent experiment.\cite{Samarth}

 In this paper, we have restricted ourselves to the process represented by a simple bubble diagram, in which only the self-energy effects are considered.
 It will be important to study the effects of vertex corrections such as ladder and skew scattering. 
 These will be reported in the future. 

{\it Note added:}
 After submitting the manuscript, we got to know that the Fermi-sea term of the Hall conductivity was calculated by Burkov\cite{Burkov2014} in essentially the same model as our ``$\bm{S}$ model''.
 However, he obtained a vanishing Hall conductivity when the chemical potential lies in the gap, which apparently disagrees with 
our result.
 This discrepancy seems to originate from the difference in the ``cut-off anisotropy'' (in our terminology).
 Namely, Burkov considers a layered system of two-dimensional continuum planes,  
which may correspond to $\lambda = \infty$ in our model (see \cref{eq:elliptic_cutoff}); 
in this special case of $\lambda = \infty$, our calculation gives $\sigma_{xy}=0$ (see \cref{fig:h_delta} or \cref{eq:g}) in agreement with Burkov's result. 
 This seems to indicate that our ``${\bm S}$ model'' can be regarded as a continuum version of the Burkov's model, and that this extension has a non-trivial consequence (Hall insulator state) due to the ultraviolet divergence. 
 We would like to thank Anton Burkov for directing our attention to Ref.~\onlinecite{Burkov2014} and for subsequent informative discussions.

\begin{acknowledgments}
 The authors would like to thank Prof.~N.~Samarth for his kind correspondence.
 This work was supported by Grants-in-Aid for Scientific Research (No.~21540336, No.~24244053 and No.~25400339) from the Japan Society for the Promotion of Science (JSPS). 
 JF is supported by JSPS Research Fellowship for Young Scientists.
\end{acknowledgments}

%
\onecolumngrid
%
\appendix
\section{Damping constants}
\label{app:gamma}
 The damping constants $\gamma_{\mu \nu} (\epsilon)$ given by \cref{eq:gamma} are calculated as follows.
 Using the relation
\begin{equation}
\begin{split}
\Im \frac{1}{D_{\bm{k}} (\epsilon + \ii 0)}
	& = - \frac{\pi}{4 \epsilon (\epsilon^2 - \epsilon_k^2) } \sum_{\eta, \zeta} \delta (\epsilon - \zeta \chi_{\bm{k}}^{\eta} (\epsilon) )
\\ & = - \frac{\pi}{ 8 \epsilon \Delta_{\bm{k}} (\epsilon)} \sum_{\eta, \zeta} \eta \delta (\epsilon - \zeta \chi_{\bm{k}}^{\eta} (\epsilon) )
,\end{split}
\label{eq:DR_Im}
\end{equation}
where
\begin{equation}
\chi_{\bm{k}}^{\eta} (\epsilon) = \sqrt{ \epsilon_k^2 + 2 \eta \Delta_{\bm{k}} (\epsilon) }
\label{eq:chi_eta}
,\end{equation}
they are written as
\begin{subequations}
\begin{align}
 \gamma_{0 0} (\epsilon)
	&= \frac{\pi}{4} n_{\mathrm{i}} u^2 \frac{1}{\epsilon} \sum_{\bm{k}, \eta, \zeta}
 	\left\{ \epsilon - \frac{\eta M \Omega}{ \Delta_{\bm{k}} (\epsilon) } \right\}
 	\delta(\epsilon - \zeta \chi_{\bm{k}}^{\eta} (\epsilon))
, \\
 \gamma_{3 0} (\epsilon)
	&= \frac{\pi}{4} n_{\mathrm{i}} u^2 \frac{1}{\epsilon} \sum_{\bm{k}, \eta, \zeta}
 	\left\{ m + \frac{\eta S \Omega}{ \Delta_{\bm{k}} (\epsilon) } \right\}
 	\delta(\epsilon - \zeta \chi_{\bm{k}}^{\eta} (\epsilon))
, \\
 \gamma_{0 z} (\epsilon)
	&= - \frac{\pi}{4} n_{\mathrm{i}} u^2 \frac{1}{\epsilon} \sum_{\bm{k}, \eta, \zeta}
	\left\{ S + \frac{\eta S k_z^2}{ \Delta_{\bm{k}} (\epsilon) } + \frac{\eta m \Omega}{ \Delta_{\bm{k}} (\epsilon) } \right\}
	\delta(\epsilon - \zeta \chi_{\bm{k}}^{\eta} (\epsilon))
,\\
 \gamma_{3 z} (\epsilon)
	&= \frac{\pi}{4} n_{\mathrm{i}} u^2 \frac{1}{\epsilon} \sum_{\bm{k}, \eta, \zeta}
	\left\{ M + \frac{\eta M k_z^2}{ \Delta_{\bm{k}} (\epsilon) } - \frac{\eta \epsilon \Omega}{ \Delta_{\bm{k}} (\epsilon) } \right\}
	\delta(\epsilon - \zeta \chi_{\bm{k}}^{\eta} (\epsilon))
.\end{align}
\label{eq:detail_gamma}
\end{subequations}
 If we define 
\begin{equation}
 \nu_{l, m}^{\eta} (\epsilon) = \sum_{\bm{k}, \zeta} \frac{k_z^{2 m}}{ [ \Delta_{\bm k} (\epsilon) ]^{l} } \delta(\epsilon - \zeta \chi_{\bm k}^{\eta}(\epsilon) )
\label{eq:gDoS_general}
,\end{equation}
the $\gamma_{\mu \nu} (\epsilon)$'s are expressed as
\begin{subequations}
\begin{align}
\gamma_{0 0} (\epsilon)
	& = \frac{\pi}{4} n_{\mathrm{i}} u^2 \frac{1}{\epsilon} \sum_{\eta}
	( \epsilon \nu_{0,0}^{\eta} (\epsilon) - \eta M \Omega \nu_{1,0}^{\eta} (\epsilon) )
\label{eq:gamma_appendix_00}
, \\
\gamma_{3 0} (\epsilon)
	& = \frac{\pi}{4} n_{\mathrm{i}} u^2 \frac{1}{\epsilon} \sum_{\eta}
	( m \nu_{0,0}^{\eta} (\epsilon) + \eta S \Omega \nu_{1, 0}^{\eta} (\epsilon) )
, \\
\gamma_{0 z} (\epsilon)
	& = - \frac{\pi}{4} n_{\mathrm{i}} u^2 \frac{1}{\epsilon} \sum_{\eta}
	( S \nu_{0,0}^{\eta} (\epsilon) + \eta S \nu_{1, 1}^{\eta} (\epsilon) + \eta m \Omega \nu_{1, 0}^{\eta} (\epsilon) )
, \\
\gamma_{3 z} (\epsilon)
	& = \frac{\pi}{4} n_{\mathrm{i}} u^2 \frac{1}{\epsilon} \sum_{\eta}
	( M \nu_{0,0}^{\eta} (\epsilon) + \eta M \nu_{1, 1}^{\eta} (\epsilon) - \eta \epsilon \Omega \nu_{1, 0}^{\eta} (\epsilon) )
\label{eq:gamma_appendix_3z}
.\end{align}
\end{subequations}
 For $(l,m) = (0,0), (1,0), (1,1)$, they are calculated as
\begin{align}
\nu_{0, 0}^{\eta} (\epsilon)
	& = \frac{|\epsilon|}{2 \pi^2} \Theta_{\eta} (\epsilon)
		\xi_{\eta}(\epsilon)
, \\
\nu_{1, 0}^{\eta} (\epsilon)
	& = \frac{|\epsilon|}{2 \pi^2} \Theta_{\eta} (\epsilon)
		h_{\eta} (\epsilon)
, \\
\nu_{1, 1}^{\eta} (\epsilon)
	& = \frac{|\epsilon|}{4 \pi^2} \Theta_{\eta} (\epsilon)
		\frac{1}{S^2 - M^2} \left(
			\xi_{\eta}(\epsilon) \sqrt{\Omega^2 + (S^2 - M^2) [\xi_{\eta}(\epsilon)]^2}
			- \Omega^2 h_{\eta} (\epsilon)
		\right)
,\end{align}
where
\begin{equation}
h_{\eta} (\epsilon) = \left\{
\begin{array}{c c}
\displaystyle
	\frac{1}{\sqrt{S^2 - M^2}} \tanh^{-1} \frac{\xi_{\eta}(\epsilon) \sqrt{S^2 - M^2}}{ \sqrt{\Omega^2 + (S^2 - M^2) [\xi_{\eta}(\epsilon)]^2 } }
&	\ \ \  (S^2-M^2 > 0)
, \\ \displaystyle
	\frac{1}{\sqrt{|S^2 - M^2|}} \tan^{-1} \frac{\xi_{\eta}(\epsilon) \sqrt{|S^2 - M^2|}}{ \sqrt{\Omega^2 - |S^2 - M^2| [\xi_{\eta}(\epsilon)]^2 } }
&	\ \ \  (S^2-M^2 < 0)
,\end{array}
\right.
\end{equation}
and $\xi_{\eta} (\epsilon)$ and $\Theta_{\eta} (\epsilon)$ are defined by \cref{eq:xi_eta,eq:Theta_eta}, respectively.

\section{``Magnetization" model}
\label{app:magnetization}
 In this Appendix, we show some details of the calculation for the ``magnetization" model, in which $\bm{M} = M \hat{z}$ and $\bm{S} = 0$ [case (i)]. 
 The denominator, $D_{\bm{k}} (\epsilon)$, of the unperturbed Green's function is given by
\begin{align}
D_{\bm{k}} (\epsilon)
	& = (\epsilon - E_{\bm{k}}^{+}) (\epsilon + E_{\bm{k}}^{+}) (\epsilon - E_{\bm{k}}^{-}) (\epsilon + E_{\bm{k}}^{-})
, \hspace{4ex}
\\ E_{\bm{k}}^{\eta}
	& = \sqrt{ k^2 + m^2 + M^2 + 2 \eta M \sqrt{ k^2_{\perp} + m^2} }
\\	& = \sqrt{ \epsilon_k^2 + 2M^2 + 2 \eta M \sqrt{ k^2_{\perp} + m^2} }
\label{eq:dispersion_M}
,\end{align}
where $\eta = \pm$, $k_{\perp}^2 = k_x^2 + k_y^2$, and $\epsilon_k^2 = k^2 + m^2 - M^2$.

 It is convenient to define the generalized density of states by 
\begin{equation}
\mathcal{N}_{l, \eta} (\epsilon) = \sum_{\bm{k}} \sum_{\zeta = \pm 1} (k_{\perp}^2 + m^2)^{l/2} \delta (\epsilon - \zeta E_{\bm{k}}^{\eta}) 
\label{eq:gDoS_M}
.\end{equation}
 Using the relation $\sum_{\zeta} \delta (\epsilon - \zeta E_{\bm{k}}^{\eta}) = 2 |\epsilon| \delta (k_z^2 - \epsilon^2 + (\sqrt{k_{\perp}^2 + m^2} + \eta M)^2)$ and performing the $\bm{k}$-integration, 
those of our interest (see below) are obtained as 
\begin{align}
\mathcal{N}_{-1, \eta} (\epsilon)
	& = \frac{|\epsilon|}{2 \pi^2} \Theta_{\eta} (\epsilon) \left( \frac{\pi}{2} - \psi_a \right)
, \\
\mathcal{N}_{0, \eta} (\epsilon)
	& = \frac{|\epsilon|}{2 \pi^2} \Theta_{\eta} (\epsilon) \left(
		|\epsilon| \cos \psi_a
		- \eta M \left( \frac{\pi}{2} - \psi_a \right) \right)
, \\
\mathcal{N}_{1, \eta} (\epsilon)
	& = \frac{|\epsilon|}{4 \pi^2} \Theta_{\eta} (\epsilon) \left(
		( m - 3 \eta M ) |\epsilon| \cos \psi_a
		+ ( \epsilon^2 + 2 M^2 ) \left( \frac{\pi}{2} - \psi_a \right) \right)
,\end{align}
where $\psi_a = \sin^{-1} [ (m + \eta M) / |\epsilon| ] $, and
\begin{equation}
\Theta_{\eta} (\epsilon) = \left\{
	\begin{array}{c c}
		1
	&	\quad ( | \epsilon | > m + \eta M)
	\\	0
	&	\quad ( \text{otherwise} )
.\end{array}
\right.
\end{equation}

 To calculate $\sigma^{(3)}_{x y}$, we note $X (\epsilon) =  - 4 m M \epsilon \left( \epsilon^2 - \epsilon_k^2  \right)$ [\cref{eq:X}] and $\partial_{\epsilon} D_{\bm{k}} =  4 \epsilon (\epsilon^2 -\epsilon_k^2 - 2M^2) $, and rewrite \cref{eq:sigma3} as
\begin{equation}
\sigma_{x y}^{(3)}
 = - \frac{ m e^2}{ 2 \mu} \sum_{\eta} \left( \eta \mathcal{N}_{-1, \eta} (\mu) + M \mathcal{N}_{-2, \eta} (\mu) \right)
 	- \frac{ m e^2}{ 2 } \sum_{\eta} \eta \sum_{\bm{k}, \zeta}
		\frac{1}{ (k_{\perp}^2 + m^2)^{3/2} }
		\int_{-\infty}^{\mu} \dd{\epsilon} \delta ( \epsilon - \zeta E_{\bm{k}}^{\eta})
\label{eq:sigma3_M_progress}
.\end{equation}
 In the second term on the right-hand side, the $\epsilon$-integral in the range $(- \infty , 0)$ vanishes because of $\sum_{\eta} \eta$ = 0, and the rest of the $\epsilon$- and $\bm{k}$-integrals are calculated as follows:
\begin{equation}
\begin{split}
\sum_{\bm{k}, \zeta}
	\frac{1}{ (k_{\perp}^2 + m^2)^{3/2} }
	\int_{0}^{\mu} \dd{\epsilon} \delta ( \epsilon - \zeta E_{\bm{k}}^{\eta})
 & = \frac{1}{(2 \pi)^2} \Theta_{\eta} (\mu) \int_{0}^{\xi_{\eta}} \dd{k_{\perp}} 
	 \frac{2 k_{\perp}}{ (k_{\perp}^2 + m^2)^{3 / 2} }
	 \int_{0}^{ \sqrt{\mu^2 - \left( \sqrt{k_{\perp}^2 + m^2} + \eta M \right)^2 } } \dd{k_z}
\\ & = \frac{\mu |\mu|}{ 2 \pi^2} \Theta_{\eta} (\mu) \int_{\psi_a}^{\psi_b} \dd{\theta}
	 \frac{ \cos^2 \theta}{  (\mu \sin \theta - \eta M)^2 }
\\ & = \frac{|\mu|}{2 \pi^2} \Theta_{\eta} (\mu)
\left[
	\frac{1}{m} \cos \psi_a
	 - \frac{1}{\mu} \int_{\psi_a}^{\psi_b} \dd{\theta}
	\left(
	 	1 + \frac{ \eta M }{ \mu \sin \theta - \eta M }
	\right)
\right]
\\ & = \frac{1}{m \mu} \left( \mathcal{N}_{0, \eta} (\mu) + \eta M \mathcal{N}_{-1, \eta} (\mu) \right)
	 - \frac{1}{\mu} ( \mathcal{N}_{-1, \eta} (\mu) + \eta M \mathcal{N}_{-2, \eta} (\mu) )
.\end{split}
\end{equation}
 Therefore, the second term in \cref{eq:sigma3_M_progress} can be rewritten as
\begin{equation}
- \frac{e^2}{2 \mu} \sum_{\eta}
\left[
\left( \eta \mathcal{N}_{0, \eta} (\mu) + M \mathcal{N}_{-1, \eta} (\mu) \right)
	 - m ( \eta \mathcal{N}_{-1, \eta} (\mu) + M \mathcal{N}_{-2, \eta} (\mu) ) 
\right]
,\end{equation}
and we obtain 
\begin{equation}
\begin{split}
\sigma_{x y}^{(3)}
	& = - \frac{e^2}{2 \mu} \sum_{\eta}
		\left( \eta \mathcal{N}_{0, \eta} (\mu) + M \mathcal{N}_{-1, \eta} (\mu) \right)
\\ & = - \sign{\mu}  \, \frac{e^2}{4 \pi^2} \sum_{\eta} \eta \,
		\Theta_{\eta} (\mu) \sqrt{\mu^2 - (m + \eta M)^2 }
.\end{split}
\label{eq:sigma3_M}
\end{equation}
 These are shown in \cref{fig:AHE} (i)  as the Fermi-sea term.

\section{``Spin" model}
\label{app:spin}
 For the ``spin" model, where $\bm{M} = 0$ and $\bm{S} = S \hat{z}$ [case (ii)], the denominator $D_{\bm{k}} (\epsilon)$ [\cref{eq:Dk_general}]  
of the unperturbed Green's function is given by
\begin{align}
D_{\bm k} (\epsilon)
	& = (\epsilon - E_{\bm{k}}^{+}) (\epsilon + E_{\bm{k}}^{+}) (\epsilon - E_{\bm{k}}^{-}) (\epsilon + E_{\bm{k}}^{-})
, \\
E_{\bm{k}}^{\eta}
	& = \sqrt{ k^2 + m^2 + S^2 + 2 \eta S \sqrt{ k^2_z + m^2} }
\\	& = \sqrt{ \epsilon_k^2 + 2 \eta S \sqrt{ k^2_z + m^2} }
\label{eq:dispersion_S}
,\end{align}
where $\epsilon_k^2 = k^2 + m^2 + S^2$.

The generalized density of states in this case is defined by
\begin{equation}
\mathcal{N}_{l, \eta} (\epsilon) = \sum_{\bm{k}, \zeta} (k_z^2 + m^2)^{l/2} \delta ( \epsilon - \zeta E_{\bm{k}}^{\eta} ) 
\label{eq:gDoS_S}
,\end{equation}
some of which are calculated as 
\begin{align}
\mathcal{N}_{-1, \eta} (\epsilon)
	& =  \frac{| \epsilon |}{2 \pi^2} \theta_{\eta}^{*}(\epsilon) \Theta_{\eta}(\epsilon)
, \\
\mathcal{N}_{0, \eta} (\epsilon)
	& = \frac{m | \epsilon |}{2 \pi^2} \sinh \theta_{\eta}^{*}(\epsilon)  \Theta_{\eta}(\epsilon)
, \\
\mathcal{N}_{1, \eta} (\epsilon)
	& = \frac{m^2 | \epsilon |}{4 \pi^2} \left( \theta_{\eta}^{*} + \sinh \theta_{\eta}^{*}(\epsilon) \cosh \theta_{\eta}^{*}(\epsilon) \right) \Theta_{\eta}(\epsilon)
,\end{align}
where $ \theta_{\eta}^{*}(\epsilon) = \cosh^{-1} [ ( |\epsilon| - \eta S ) / m ]$ and
\begin{equation}
\Theta_{\eta} (\epsilon) = \left\{
	\begin{array}{c c}
		1
	&	\quad ( | \epsilon | > m + \eta S)
	\\	0 
	&	\quad ( \text{otherwise} )
.\end{array}
\right.
\end{equation}

 To calculate $\sigma^{(3)}_{x y}$, we first express it as 
\begin{equation}
\sigma_{x y}^{(3)}
 = - \frac{e^2}{2 \mu^2} \sum_{\eta} \left( S \mathcal{N}_{0, \eta} (\mu) + \eta \mathcal{N}_{1, \eta} (\mu) \right)
 - \frac{e^2}{2} \sum_{\bm{k}, \eta, \zeta} \eta \frac{ \sqrt{k_z^2 + m^2} + \eta S }{\zeta \left( E_{\bm{k}}^{\eta} \right)^3 }
	\int_{-\infty}^{\mu} \dd{\epsilon} \delta(\epsilon - \zeta E_{\bm{k}}^{\eta})
,\end{equation}
by noting that $ X(\epsilon) = - S \{ D_k + 4 (k_z^2 + m^2) ( \epsilon^2 - \epsilon_k^2 + 2 S^2 ) \} $ and $\partial_{\epsilon} D_k = 4 \epsilon (\epsilon^2 - \epsilon_k^2) $. 
 In the second term on the right-hand side, the $\epsilon$-integral in the range $(-\infty, 0)$ is calculated as 
\begin{equation}
 - \frac{e^2}{2} \sum_{\bm{k}, \eta, \zeta} \eta \frac{ \sqrt{k_z^2 + m^2} + \eta S }{\zeta \left( E_{\bm{k}}^{\eta} \right)^3 }
	\int_{-\infty}^{0} \dd{\epsilon} \delta(\epsilon - \zeta E_{\bm{k}}^{\eta})
=	\frac{e^2}{2} \sum_{\bm{k}, \eta} \frac{S + \eta \sqrt{k_z^2 + m^2}}{ \left( E_{\bm{k}}^{\eta} \right)^3 }
.\end{equation}
 In calculating this $\bm{k}$-integral, a care is needed since it is ultraviolet divergent. 
 To manage this divergence, we introduce a momentum cut-off $\Lambda$ and limit the integration to a spherical region, $|{\bm k}| < \Lambda$, and then let $\Lambda \to \infty$. 
 This proceeds as follows; we first write 
\begin{equation}
\begin{split}
& \frac{e^2}{2 (2 \pi)^2}
	\lim_{\Lambda \to \infty}
	\sum_{\eta} \int_{0}^{\Lambda} \dd{k_z} (S + \eta \sqrt{k_z^2 + m^2})
	\int_{0}^{\sqrt{\Lambda^2 - k_z^2}} \dd{k_{\perp}} 2 k_{\perp} \left( \frac{1}{ k_{\perp}^2 + (S + \eta \sqrt{k_z^2 + m^2})^2} \right)^{3/2}
\\ & = - \frac{e^2}{(2 \pi)^2}
	\lim_{\Lambda \to \infty}
	\sum_{\eta} \int_{0}^{\Lambda} \dd{k_z}
	\frac{S + \eta \sqrt{k_z^2 + m^2} }{ \sqrt{ \Lambda^2 - k_z^2 + \left( S + \eta \sqrt{k_z^2 + m^2} \right)^2 } }
,\end{split}
\end{equation}
then scale as $k_z = \Lambda t, \tilde{m} = m / \Lambda, \tilde{S} = S / \Lambda$, and sum over $\eta$. 
 The result is the left-hand side of 
\begin{equation}
 - \frac{S e^2}{( 2 \pi)^2}
\lim_{\Lambda \to \infty}
	\int_0^{1} \dd{t}
		\left\{
			\mathcal{A} (t) - 4 \mathcal{B} (t)
		\right\}
= - \frac{S e^2}{ 3 \pi^2} 
\label{eq:calc_spin_1}
,\end{equation}
with 
\begin{align}
\mathcal{A} (t)
	& = \sum_{\eta} \left\{ 1 - t^2 + \left( \tilde{S} + \eta \sqrt{t^2 + \tilde{m}^2} \right)^2 \right\}^{-1/2}
, \\
\mathcal{B} (t)
	& = \frac{ t^2 + \tilde{m}^2}{  \displaystyle \sum_{\eta} \left\{ 1 - t^2 + \left( \tilde{S} + \eta \sqrt{ t^2 + \tilde{m}^2 } \right)^2 \right\}^{1/2}
        \displaystyle \prod_{\eta'} \left\{ 1 - t^2 + \left( \tilde{S} + \eta' \sqrt{ t^2 + \tilde{m}^2 } \right)^2 \right\}^{1/2} } 
.\end{align}
 We then let $\Lambda \to \infty$, thus $\tilde{S} \to 0, \tilde{m} \to 0$, $\mathcal{A} \to 2$ and $\mathcal{B} \to t^2/2$, and integrate over $t$; this leads to the right-hand side of \cref{eq:calc_spin_1}.
 The rest of the integral, which is divergence-free, is calculated as follows, 
\begin{equation}
\begin{split}
& - \frac{e^2}{2 (2 \pi)^2} \sum_{\eta} \Theta_{\eta} (\mu) \int_{0}^{\xi_{\eta}} \dd{k_z} (S + \eta \sqrt{k_z^2 + m^2})
	\int_{0}^{\alpha_{\eta}} \dd{t} \left( \frac{1}{ t + (S + \eta \sqrt{k_z^2 + m^2})^2} \right)^{3/2}
\\ & = \frac{e^2}{2 |\mu|} \frac{1}{2 \pi^2} \sum_{\eta} \Theta_{\eta} (\mu) \int_{0}^{\xi_{\eta}} \dd{k_z}
	 \left( S - \eta |\mu| + \eta \sqrt{k_z^2 + m^2} \right)
\\ & = \frac{e^2}{2 |\mu|} \frac{1}{2 \pi^2} \sum_{\eta} \Theta_{\eta} (\mu)
	\int_{0}^{\theta^{*}_{\eta}} \dd{\theta} m \cosh \theta
	 \left( S - \eta |\mu| + \eta m \cosh \theta \right)
\\ & = \frac{e^2}{2 \mu^2} \sum_{\eta} \left\{ (S - \eta |\mu|) \mathcal{N}_{0, \eta} (\mu) + \eta \mathcal{N}_{1, \eta} (\mu) \right\}
.\end{split}
\end{equation}
 Therefore we obtain
\begin{equation}
\begin{split}
\sigma_{x y}^{(3)}
	& = - \frac{S e^2}{3 \pi^2} - \frac{e^2}{2 |\mu| } \sum_{\eta} \eta \mathcal{N}_{0, \eta} (\mu)
\\ & = - \frac{S e^2}{3 \pi^2} - \frac{e^2}{4 \pi^2} \sum_{\eta} \eta \Theta_{\eta} (\mu) \sqrt{(| \mu | - \eta S)^2 - m^2}
.\end{split}
\label{eq:sigma^3_xy_S_result}
\end{equation}
This is plotted in \cref{fig:AHE} (ii) as the Fermi-sea term. 
 The first term corresponds to $\sigma_{xy} (\mu=0)$. 

 It should be noted that, if another cut-off scheme is adopted, one would obtain a different result in general.  
 For example, if we introduce two cut-off parameters, $\Lambda_{\perp}$ and $\Lambda_{z}$, limit the integration to a cylindrical region, $|{\bm k}_\perp| < \Lambda_{\perp}$ and $|k_z| < \Lambda_{z}$, and let $\Lambda_{\perp} \to \infty$ and $\Lambda_{z} \to \infty$ by keeping the ratio $\Lambda_{\perp} / \Lambda_z \equiv r$ constant, the first term of \cref{eq:sigma^3_xy_S_result} is replaced by 
\begin{equation}
 \sigma_{xy} (\mu=0) =- \frac{ S e^2}{2 \pi^2 \sqrt{r^2 + 1 }}
\label{eq:s_xy_N}
.\end{equation}
 Another, probably more natural  cut-off which allows anisotropy would be an ``elliptical'' cut-off [\cref{eq:elliptic_cutoff}], 
\begin{equation}
 k_{\perp}^2 + (1 + \lambda) k_z^2 \le \Lambda^2
, \qquad  (-1 < \lambda < \infty)
\label{eq:elliptic_cutoff_app}
.\end{equation}
 In this case, the first term of \cref{eq:sigma^3_xy_S_result} is replaced by
\begin{equation}
 \sigma_{xy} (\mu=0) = - \frac{S e^2}{3 \pi^2} \, g(\lambda)
,\end{equation}
where 
\begin{equation}
 g (\lambda) =
\left\{
	\begin{array}{c c}
	\displaystyle
		\frac{3}{2\lambda }
		\left[
			(1+\lambda) \frac{\tanh^{-1} \sqrt{-\lambda}}{\sqrt{-\lambda}} - 1 \,
		\right]
	& \qquad  ( - 1 < \lambda < 0)
	, \\ \displaystyle
		\frac{3}{ 2\lambda }
		\left[
			(1+\lambda) \frac{\tan^{-1} \sqrt{\lambda}}{\sqrt{\lambda}} - 1 \, 
		\right]
	& \qquad ( \lambda > 0)
,\end{array}
\right.
\label{eq:g}
\end{equation}
which is plotted in \cref{fig:h_delta}. 
 The value $ g (-1) = 3/2 $ at $\lambda = -1$ is consistent with \cref{eq:s_xy_N} with $r = 0$. 

\section{Coexistent' Model}
\label{app:coexistent}
 For the ``coexistent'' model with $\bm{M} = \bm{S} = S \hat{z}$ [case (iii)], the denominator  $D_{\bm{k}} (\epsilon)$ [\cref{eq:Dk_general}] of the Green's function is given by
\begin{align}
D_k (\epsilon)
	& = (\epsilon - S - E_{k}^+) (\epsilon + S - E_{k}^+) (\epsilon - S + E_{k}^-) (\epsilon + S + E_{k}^-)
\\	& = \prod_{\eta, \zeta = \pm} (\epsilon -  \eta S - \zeta E_{k}^{\eta})
, \\
E_{k}^{\eta}
	& = \sqrt{k^2 + (m + \eta  S)^2 }
\label{eq:dispersion_MS}
.\end{align}
 In this case, the electron dispersion $\epsilon = \eta S + \zeta E_k^{\eta}$ is isotropic but there is a loss of particle-hole symmetry, in contrast to the cases (i) and (ii). 

The generalized density of states is defined by
\begin{equation}
\begin{split}
\mathcal{N}_{n,l}^{\eta} (\epsilon)
	& \equiv \sum_{\bm{k}} \sum_{\zeta} \frac{(k_{\perp}^2 - k_z^2)^l}{(\epsilon^2 - \epsilon_k^2)^n}
		\delta (\epsilon - \eta S - \zeta E_k^{\eta})
\\ & = \sum_{\bm{k}} \sum_{\zeta} \frac{(k^2 / 3)^{l}}{(\epsilon^2 - \epsilon_k^2)^n}
		\delta (\epsilon - \eta S - \zeta E_k^{\eta}) , \quad \quad (l = 0,1) 
.\end{split}
\label{eq:gDoS_MS}
\end{equation}
 In the second equality, which holds for $l = 0$ and 1, we have noted that the energy dispersion is isotropic. 
 Explicitly, they are evaluated as 
\begin{equation}
\begin{split}
\mathcal{N}_{n,l}^{\eta} (\epsilon) 
	& = \frac{1}{2 \pi^2} \frac{1}{3^l} \frac{| \epsilon - \eta S |}{ \{ 2 \eta S (\epsilon + m) \}^n}
		\xi_{\eta}^{2l + 1} \Theta_{\eta} (\epsilon)
, \qquad (l = 0,1) 
,\end{split}
\end{equation}
with $\xi_{\eta} = \sqrt{ (\epsilon + m) (\epsilon - m - 2 \eta S) }$ and
\begin{equation}
\Theta_{\eta} (\epsilon) = \left\{
	\begin{array}{c c}
		1
	&	\qquad (\epsilon < - m, \, \epsilon > m + 2 \eta S) 
	\\	0
	&	\qquad (\text{otherwise}).
\end{array}
\right.
\end{equation}

 To calculate $\sigma^{(3)}_{x y}$, we note $ X (\epsilon) = S D_k - 2 S (\epsilon^2 - \epsilon_k^2) \{ (\epsilon + m)^2 - k^2 + 2 k_z^2 \}$ and $\partial_{\epsilon} D_k = 4 \{ \epsilon ( \epsilon^2 - \epsilon_k^2) - 2 S^2 (\epsilon + m) \}$, and express it as
\begin{equation}
\begin{split}
\sigma_{x y}^{(3)}
	& = - \frac{e^2}{6} \sum_{\eta} \frac{ \mu + 2 m + \eta S }{ (\mu - \eta S)^2 }
		\eta \mathcal{N}^{\eta}_{0, 0} (\mu)
\\ & \hspace{1em}
		- \frac{e^2}{2} \sum_{\bm{k}, \eta, \zeta}
		\frac{\eta}{k^2} \left[
			(m + \eta S) \frac{(k_{\perp}^2 - 2 k_z^2)}{(\zeta E_k^{\eta})^3}
			- \frac{2 k_z^2}{ k^2 } \left( \frac{m + \eta S}{\zeta E_k^{\eta}} \right)^3
			+ \frac{2 k_z^2}{ k^2 }
		\right]
		\int_{-\infty}^{\mu} \dd{\epsilon} \delta (\epsilon - \eta S - \zeta E_k^{\eta})
,\end{split}
\label{eq:sigma3_coex}
\end{equation}
where we used
\begin{align}
 \partial_{\epsilon} D_k  
	& = 8 \eta \zeta S E_k^{\eta} (\epsilon + m)  
, \\
\epsilon^2 - \epsilon_k^2  
	& = 2 \eta S ( \epsilon + m )  
, \\
\frac{1}{\epsilon + m}  
	& = - \frac{1}{k^2} ( \epsilon + m - 2 \zeta E_k^{\eta} )  
,\end{align}
which hold under the presence of $\delta (\epsilon - \eta S - \zeta E_k^{\eta})$. 
 In the second term on the right-hand side of \cref{eq:sigma3_coex}, the $\epsilon$-integral from the range $(- \infty,0)$ is divided into the convergent part ($\mathcal{W}_0$) and the conditionally-convergent part ($\mathcal{W}_1$) due to ultraviolet divergence. 
 They are respectively evaluated as follows;  For $\mathcal{W}_0$, 
\begin{equation}
\begin{split}
\mathcal{W}_0
	& \equiv - \frac{e^2}{2} \sum_{\bm{k}, \eta, \zeta} \frac{\eta}{k^2}
	\left[
		- \frac{2 k_z^2}{ k^2 } \left( \frac{m + \eta S}{\zeta E_k^{\eta}} \right)^3
		+ \frac{2 k_z^2}{ k^2 }
	\right]
	\int_{-\infty}^{0} \dd{\epsilon} \delta (\epsilon - \eta S - \zeta E_k^{\eta})
\\ & = - \frac{e^2}{3} \sum_{\bm{k}, \eta}
	\frac{\eta}{k^2} \frac{(m + \eta S)^3}{ \left( E_k^{\eta} \right)^3 }
\\ & = - \frac{e^2}{3} \frac{1}{2 \pi^2} \sum_{\eta} \eta
	\int_0^{\infty} \dd{k} \frac{(m + \eta S)^3}{ \left( E_k^{\eta} \right)^3 }
\\ & = - \frac{S e^2}{3 \pi^2}  
.\end{split}
\end{equation}
 For $\mathcal{W}_1$, we introduce a cut-off as $k_{\perp}^2 + ( 1 + \lambda) k_z^2 \le \Lambda^2$ [\cref{eq:elliptic_cutoff} or \cref{eq:elliptic_cutoff_app}], and evaluate as 
\begin{equation}
\begin{split}
\mathcal{W}_1
	& \equiv - \frac{e^2}{2} \sum_{\bm{k}, \eta, \zeta}
		\frac{\eta \zeta}{(E_k^{\eta})^3} \frac{1}{k^2} (m + \eta S) (k_{\perp}^2 - 2 k_z^2)
		\int_{-\infty}^{0} \dd{\epsilon} \delta (\epsilon - \eta S - \zeta E_k^{\eta})
\\ & = \lim_{\Lambda \to \infty}
	\frac{e^2}{2 ( 2 \pi)^2} \sum_{\eta} \eta ( m + \eta S)
	\int_0^{1/\sqrt{1+\lambda}} \dd{q}
	\int_{q^2}^{1 - \lambda q^2} \dd{p}
	\frac{1}{( p + \epsilon_{\eta}^2 )^{3/2}} \frac{p - 3 q^2}{p}
\\ & = - \lim_{\epsilon_{\eta} \to 0}
	\frac{e^2}{( 2 \pi)^2} \sum_{\eta} \eta ( m + \eta S)
	\Bigl( \mathcal{C}_{\eta} (\lambda) + \mathcal{D}_{\eta} (\lambda) \Bigr)
,\end{split}
\end{equation}
where we put $k_{\perp}^2 = \Lambda^2 (p - q^2),\, k_z = \Lambda q$, $\epsilon_{\eta} = (m + \eta S) / \Lambda$, and
\begin{align}
\mathcal{C}_{\eta} (\lambda)
	& = \int_0^{1/\sqrt{1+\lambda}} \dd{q}
	\left(
		\frac{1}{ \sqrt{1 - \lambda q^2 + \epsilon_{\eta}^2} }
		\left( 1 + \frac{3 q^2}{\epsilon_{\eta}^2} \right)
	- \frac{3 q^2}{2 \epsilon_{\eta}^3}
		\log \frac{ \sqrt{1 - \lambda q^2+\epsilon_{\eta}^2} + \epsilon_{\eta} }
			{ \sqrt{1 - \lambda q^2+\epsilon_{\eta}^2} - \epsilon_{\eta} }
	\right)
, \\
\mathcal{D}_{\eta} (\lambda)
	& = \int_0^{1/\sqrt{1+\lambda}} \dd{q}
	 \left( 
		\frac{3 q^2}{2 \epsilon_{\eta}^3}
		\log \frac{ \sqrt{q^2+\epsilon_{\eta}^2} + \epsilon_{\eta} }
			{ \sqrt{q^2+\epsilon_{\eta}^2} - \epsilon_{\eta} }
		- \frac{1}{ \sqrt{ q^2 + \epsilon_{\eta}^2} }
		\left( 1 + \frac{3 q^2}{\epsilon_{\eta}^2} \right)
	\right)
.\end{align}
 The $q$-integration and the limit $\epsilon_{\eta} \to 0$ are mutually commutative in $\mathcal{C}_{\eta} (\lambda)$, hence we can take the limit $\epsilon_{\eta} \to 0$ first, 
\begin{equation}
\lim_{\epsilon_{\eta} \to 0} \mathcal{C}_{\eta} (\lambda)
 = \int_0^{1/\sqrt{1+\lambda}} \dd{q}
	\frac{1}{\sqrt{1 - \lambda q^2}}
	\left( 1 - \frac{q^2}{1 - \lambda q^2} \right)
 = \frac{2}{3} g(\lambda)
,\end{equation}
where $g(\lambda)$ is given by \cref{eq:g}.
 For $\mathcal{D}_{\eta} (\lambda)$, they do not commute and we have to perform the $q$-integral first 
and then take the limit $\epsilon_{\eta} \to 0$,
\begin{equation}
\begin{split}
\lim_{\epsilon_{\eta} \to 0} \mathcal{D}_{\eta} (\lambda) & = 
	\lim_{\epsilon_{\eta} \to 0}
	\int_0^{1/\sqrt{1+\lambda}} \dd{q}
	 \left( 
		\frac{3 q^2}{\epsilon_{\eta}^3}
		\log \frac{ \sqrt{q^2+\epsilon_{\eta}^2} + \epsilon_{\eta} }{ q }
		- \frac{3}{\epsilon_{\eta}^2} \sqrt{ q^2 + \epsilon_{\eta}^2} + \frac{2}{ \sqrt{ q^2 + \epsilon_{\eta}^2} }
	\right)
\\ & =
	\lim_{\epsilon_{\eta} \to 0}
	\left[
		\frac{1}{2 \epsilon_{\eta}^3}
		\left\{
			q \epsilon_{\eta} \sqrt{q^2 + \epsilon_{\eta}^2}
			+ 2 q^3 \log \frac{\sqrt{q^2 + \epsilon_{\eta}^2} + \epsilon_{\eta}}{ q }
			- \epsilon_{\eta}^3 \log \left( q + \sqrt{q^2 + \epsilon_{\eta}^2} \right)
		\right\}
\right. \\ & \hspace{4em} \left.
		- \frac{3}{2 \epsilon_{\eta}^2} \left\{ q \sqrt{ q^2 + \epsilon_{\eta}^2} + \epsilon_{\eta}^2 \log ( q + \sqrt{ q^2 + \epsilon_{\eta}^2} ) \right\}
		+ 2 \log ( q + \sqrt{ q^2 + \epsilon_{\eta}^2} )
	\right]_{q=0}^{q =1/\sqrt{1+\lambda}}
\\ & = - \frac{2}{3}
.\end{split}
\end{equation}
 Consequently, we obtain $\mathcal{W}_1 = - (S e^2/3 \pi^2) ( g(\lambda) - 1 )$.
 Note that $\mathcal{W}_1 (\lambda=0) = 0$ for the isotropic cut-off, $k^2 < \Lambda$.
 For the rest part of the $\epsilon$-integral of \cref{eq:sigma3_coex}, since there is no divergence and the dispersion is isotropic, we evaluate it by replacing $k_\perp^2$ by $2k^2/3$ and $k_z^2$ by $k^2/3$ as 
\begin{equation}
\begin{split}
\frac{e^2}{3} \sum_{\bm{k}, \eta, \zeta} \frac{\eta}{k^2}
\left[
	\left( \frac{m + \eta S}{ \zeta E_k^{\eta}} \right)^3 - 1
\right]
\int_{0}^{\mu} \dd{\epsilon} \delta (\epsilon - \eta S - \zeta E_k^{\eta})
	& = \frac{e^2}{3} \sum_{\bm{k}, \eta} \Theta_{\eta} (\mu)
	\frac{\eta}{k^2} \left[ \mathrm{sign} (\mu) \left( \frac{m + \eta S}{E_k^{\eta}} \right)^3 - 1 \right]
\\ & = \frac{e^2}{3} \frac{1}{2 \pi^2} \sum_{\eta} \eta \Theta_{\eta} (\mu)
 	 \int_0^{\xi_{\eta}} \dd{k} \left[ \mathrm{sign} (\mu) \left( \frac{m + \eta S}{E_k^{\eta}} \right)^3 - 1 \right]
\\ & = \frac{e^2}{3} \frac{1}{2 \pi^2} \sum_{\eta} \eta \Theta_{\eta} (\mu)
 	 \left[ \mathrm{sign} (\mu) (m + \eta S)  \tanh \theta_{\eta}^{*} - \xi_{\eta} \right]
\\ & = - \frac{e^2}{3} \sign{\mu} \sum_{\eta}
	\frac{ \mu - m - 2 \eta S } {( \mu - \eta S)^2}
	\eta \mathcal{N}_{0, 0}^{\eta} (\mu)
,\end{split}
\end{equation}
where $\sinh \theta_{\eta}^{*} = \xi_{\eta} / (m + \eta S)$ because of $\Theta_{\eta} (\mu)$.
 Here we have used $\sign{\mu} = \sign{\mu - \eta S}$.
 Therefore, we obtain
\begin{equation}
\begin{split}
\sigma^{(3)}_{x y}
& = - \frac{S e^2}{3 \pi^2} \, g(\lambda) 
- \frac{e^2}{6} \sum_{\eta} \eta \mathcal{N}_{0, 0}^{\eta} (\mu)
	\frac{ 2 \, \mathrm{sign} (\mu) ( \mu - m - 2 \eta S) + \mu + 2 m + \eta S } {( \mu - \eta S)^2}
\\ & = - \frac{S e^2}{3 \pi^2} \, g(\lambda) 
- \frac{e^2}{12 \pi^2} \sum_{\eta} \eta \, \Theta_{\eta} (\mu) \xi_{\eta}
		\frac{ 2 \, \mathrm{sign} (\mu) ( \mu - m - 2 \eta S ) + \mu + 2 m + \eta S } {| \mu - \eta S |}
.\end{split}
\end{equation}
 This is plotted in \cref{fig:AHE} (iii) as the Fermi-sea term for $\lambda=0$. 
\section{Symmetry Relations}
\label{app:sym}
 The relations in \cref{eq:symmetry_xy} are derived as follows. 
 To be explicit, let us consider $Q_{ij} (\ii \omega_{\lambda})$ given by \cref{eq:Q}. 
 We insert $1 = U U^\dagger$ in all four interspaces in the trace, with $U=\rho_1$. 
 Under this `unitary transformation', the velocity matrix does not change, $U^\dagger {\bm v} U = {\bm v}$, while the Green's function changes to 
\begin{equation}
 U^\dagger \tilde G_{\bm{k}} (\ii \epsilon_n ; m, M, S, \mu)U
	= \tilde G_{\bm{k}} (\ii \epsilon_n ; -m, -M, S, \mu) 
	= - \tilde G_{\bm{k}} (-\ii \epsilon_n ; m, M, -S, -\mu)
\label{eq:UGU}
.\end{equation}
 Note that the Green's function $\tilde G_{\bm{k}} (\ii \epsilon_n)$ [\cref{eq:G_tilde}] explicitly includes the chemical potential $\mu$. 
 Note also that \cref{eq:UGU} holds even in the presence of self-energy, \cref{eq:self-energy}, as can be seen from Feynman diagrams. 
 By changing the variable as $\ii \epsilon_n  \to  \ii \omega_{\lambda} - \ii \epsilon_{n}$ and using the cyclic property of the trace, one can show that 
\begin{equation}
 Q_{ij} (\ii \omega_{\lambda}; \mu, M, S) = Q_{ji} (\ii \omega_{\lambda}; -\mu, M, -S)
.\end{equation}
 This leads to 
\begin{equation}
 \sigma _{ij} ( \mu , M, S)  = \sigma_{ji} (-\mu , M, -S) 
\label{eq:symmetry_ij}
\end{equation}
from \cref{eq:conductivity}. 
 Since $\sigma_{xy}$ is the anti-symmetric part of  $\sigma_{ij}$, we obtain $\sigma _{xy} (\mu , M, S) = - \sigma_{x y} (-\mu , M, -S)$. 
 This is the first equality in \cref{eq:symmetry_xy}. 
 The second equality in \cref{eq:symmetry_xy} follows if we note that $\sigma_{xy}$ changes sign under $(M,S) \to (-M, -S)$. 
 This can be shown in a similar way by taking $U = \sigma^y$ and changing variables as  $(k_x, k_y, k_z) \to (-k_x, k_y, -k_z)$. 
\twocolumngrid
%


\begin{thebibliography}{99}
\bibitem{com5} Mathematically, under the Lorentz transformation, 
${\bm M}$ transforms as spatial components of an anti-symmetric tensor 
$\sigma^{\mu\nu} = [\gamma^\mu , \gamma^\nu]/2i$, where $\gamma^\mu$'s are  Dirac gamma matrices, 
and can couple to an electromagnetic field $F_{\mu\nu}$. 
 On the other hand, ${\bm S}$ transforms as spatial components of a pseudo-vector $\gamma^5 \gamma^\mu$, 
but there is no physical pseudo-vector fields.
\bibitem{com6} Note, however, that if one integrates out electrons (and positrons) and look at the effective free energy, there is a coupling between ${\bm S}$ and the magnetic field for finite $\mu$.
\bibitem{com2} 
 Consider a system of interacting Dirac particles and applying to it a mean-field approximation. 
 It is true that when the ferromagnetism arises from the Hartree-type mean field in the dipole-dipole interaction, which is mediated by real magnetic field, the order parameter should be ${\bm M}$. 
 However, when the ferromagnetism arises, as in the non-relativistic case, from the Fock-type mean field in the Coulomb interaction (or any other interactions), there is no reason to prefer ${\bm M}$ over ${\bm S}$.
\bibitem{com_elmag_eff} 
 In the ``effective'' Dirac model, while the electromagnetic minimal coupling leads to a coupling, $- {\bm M} \!\cdot\! {\bm B}$, between ${\bm M}$ and the magnetic field $ {\bm B}$, there can be other couplings due to non-minimal coupling. 
 For example, if $\bm{\sigma}$ corresponds to the real spin and is associated with a magnetic moment $g \mu_{\rm B} {\bm \sigma}$ (instead of  $g \mu_{\rm B} {\bm \sigma} \rho_3$), a direct Zeeman coupling $- g \mu_{\rm B}{\bm \sigma} \!\cdot\! {\bm B} \sim - {\bm S} \!\cdot\! {\bm B}$ is expected between ${\bm S}$ and ${\bm B}$.
\bibitem{com_calc} In the derivation, we noted that 
 $D'' \equiv {\rm Im} D_{\bm k}(\epsilon + i\eta) = \eta (\partial_\epsilon D_{\bm k})$ 
for $\eta \to +0$, and 
(ii)  $\sign {D''}= \sign {\partial_\epsilon D_{\bm k}}$. 
 From (ii), we have $\sign {D''} |\partial_\epsilon D_{\bm k}| = \partial_\epsilon D_{\bm k}$. 
 From (i), we neglected the term containing 
$\partial_\epsilon \sign {D''} \propto \delta (\partial_\epsilon D_{\bm k})$. 
 This is because this term then contains $\delta (\partial_\epsilon D_{\bm k}) \delta (D_{\bm k}) $, 
which vanish if the equation $D(\epsilon)=0$ does not have multiple root(s) as in the present case. (Note that there are no band crossings for the parameters we use in this paper.)
\bibitem{com_loose_notation}
 \Cref{eq:Pauli} is loosely written by mixing real and ${\bm k}$-spaces, and as if $V_{\rm imp}$ conserves momentum, to avoid notational complications which are not essential here.
\end{thebibliography}
\end{document}